\documentclass[]{aastex631}

\usepackage{amsmath}
\usepackage{graphicx}
\usepackage{subfigure}
\usepackage{ulem}
\usepackage{threeparttable}

\begin{document}

\title{ALMA observations of Magnetic Fields in the Massive Star-forming Region IRAS 18360-0537}

\author[0000-0002-5845-9397]{Shixian Mo}
\affiliation{School of Astronomy and Space Science, Nanjing University, 163 Xianlin Avenue, Nanjing 210023}
\affiliation{Key Laboratory of Modern Astronomy and Astrophysics (Nanjing University), Ministry of Education, Nanjing 210023}

\author[0000-0002-5093-5088]{Keping Qiu}
\affiliation{School of Astronomy and Space Science, Nanjing University, 163 Xianlin Avenue, Nanjing 210023}
\affiliation{Key Laboratory of Modern Astronomy and Astrophysics (Nanjing University), Ministry of Education, Nanjing 210023}

\author[0000-0003-2384-6589]{Qizhou Zhang}
\affiliation{ Center for Astrophysics |Harvard \& Smithsonian, 60 Garden Street, Cambridge, MA 02138, USA}

\author[0000-0002-4774-2998]{Junhao Liu}
\affiliation{School of Astronomy and Space Science, Nanjing University, 163 Xianlin Avenue, Nanjing 210023}
\affiliation{Key Laboratory of Modern Astronomy and Astrophysics (Nanjing University), Ministry of Education, Nanjing 210023}

\author[0000-0002-3829-5591]{Josep Miquel Girart}
\affiliation{Institut de Ciències de l’Espai (ICE, CSIC), Can Magrans s/n, 08193, Cerdanyola del Vallés, Catalonia, Spain}
\affiliation{Institut d’Estudis Espacials de Catalunya (IEEC), 08034, Barcelona, Catalonia, Spain}

\author[0000-0003-2300-2626]{Hauyu Baobab Liu}
\affiliation{Department of Physics, National Sun Yat-Sen University, No. 70, Lien-Hai Road, Kaohsiung City 80424}
\affiliation{Center of Astronomy and Gravitation, National Taiwan Normal University, Taipei 116}

\author[0000-0002-7402-6487]{Zhi-Yun Li}
\affiliation{Astronomy Department, University of Virginia, Charlottesville, VA 22904-4325, USA}

\author[0000-0003-1275-5251]{Shanghuo Li}
\affiliation{School of Astronomy and Space Science, Nanjing University, 163 Xianlin Avenue, Nanjing 210023}
\affiliation{Key Laboratory of Modern Astronomy and Astrophysics (Nanjing University), Ministry of Education, Nanjing 210023}

\author[0000-0002-9774-1846]{Huei-Ru Vivien Chen}
\affiliation{Institute of Astronomy and Department of Physics, National Tsing Hua University, Hsinchu 300044}

\correspondingauthor{Keping, Qiu}
\email{kpqiu@nju.edu.cn}

\begin{abstract}
Assessing the significance of magnetic fields in high-mass star formation remains one of the most challenging topics in astrophysics. In this study, we present full polarization observations obtained from the Atacama Large Millimeter/Submillimeter Array (ALMA) of the high-mass star-forming region IRAS18360-0537. The polarized dust emission at 1.3 mm reveals a clear hourglass-shaped morphology of the magnetic field. 
Interestingly, the magnetic field orientation is nearly perpendicular to both the outflow and core rotation axes, while it aligns with the elongation of the core. This orientation poses challenges for interpretation, particularly in light of the strong magnetic field strength estimated using the Davis-Chandrasekhar-Fermi method. Several scenarios provide insights into the underlying reasons for this magnetic field morphology. A clear velocity gradient seen in high-density tracing of molecular spectral lines indicates that the core is fast-rotating. The curved outskirts of the magnetic fields coincide with the outflow cavity, suggesting a possible influence from the outflow. The accretion flows along the core's elongation are also notable. Our study shows that the morphology of the magnetic field is probably highly influenced by the gas bulk motions.
\end{abstract}


\section{Introduction} \label{sec:intro}

High-mass stars play a fundamental role in the chemical enrichment of the interstellar medium and the evolution of galaxies. However, our understanding of the physical mechanisms controlling high-mass star formation remains limited. The consensus is that gravity, magnetic fields, and turbulence are key factors in this complex, multi-scale process. One of the central debates in the literature concerns the relative importance of magnetic fields compared to turbulence and gravity. Regarding magnetic fields, there exist strong magnetic field theories (e.g., \citealp{Mouschovias1976note, Shu1987star, Mouschovias1999magnetic}) and weak magnetic field theories (e.g., \citealp{Elmegreen2000obervation, Maclow2004control, Elmegreen2004interstellar}). In the strong magnetic field theories, the magnetic fields dominate and become the primary factor in resisting gravitational contraction. Meanwhile, the cores accrete gas through the gravity-driven ambipolar diffusion. Eventually, the collapse occurs once the mass-to-flux ratio exceeds the critical value (e.g., \citealp{Mestel1956star, Myers1988magnetic}). While the weak magnetic field theories suggest that formation processes are dominated by supersonic turbulence and clumps and cores are formed through the super-Alfv\'{e}nic turbulent compression at the intersection of supersonic shocks (e.g., \citealp{Ballesteros-Paredes2003dynamic, Maclow2004control, Klessen2005quiescent}). Although the starting points of the two theories are different, they cannot be easily distinguished, because turbulent, kinetic energies, and magnetic energies have the same order of magnitude from the observations of the molecular clouds \citep{Crutcher1999magnetic}. Several observations and numerical studies provide support for strong magnetic field models (e.g., \citealp{Mouschovias1976nonhomologous1, Mouschovias1976nonhomologous2, Galli1993collapse1, Galli1993collapse2, Allen2003collapse, Girart2006magnetic, Rao2009iras, Girart2009magnetic, Stephens2013magnetic, Qiu2014submillimeter, Maury2018magnetically, Kwon2019highly, Cortes2021magnetic}), while numerous studies also advocate for weak magnetic field models (e.g., \citealp{Padoan1999superalfvenic, Padoan2002thestellar, Maclow2004control, Crutcher2009testing, Federrath2010comparing, Tan2013thedynamics, Hull2017alma, Wang2020collapsing, Eswaraiah2021revealing, Karoly2023studying, Rawat2024understanding}) in both low- and high-mass star-forming regions. 

One of the approaches to determining whether magnetic fields are dominant is based on their morphology. The dust polarized emission effectively traces the plane-of-sky magnetic field in molecular clouds. The underlying concept behind this method is elucidated by the Radiative Alignment Torque (RAT) theory, which states that the dust grains with sizes smaller than 100$\mu$m tend to partially orient their short axes in parallel to the local magnetic field due to the radiative torques of anisotropic radiation fields (\citealp{Lazarian2007radiative, Hoang2009grain, Andersson2015interstellar}). Thanks to the advancement in (sub)mm interferometers over the past decades, morphologies of magnetic fields in high-mass star-forming regions, which are relatively further away from the Sun, have been probed through polarized dust emission across various physical scales. Among these magnetic field structures, the ordered hourglass-shaped fields are often regarded as consistent with the predictions of strong magnetic field theories (\citealp{Shu1987star, Basu1994magnetic, Mouschovias2006observational}). These hourglass fields are typically interpreted as indicative of strong coupling between the magnetic fields and dense gas, as well as the presence of ambipolar diffusion (e.g., \citealp{Galli1993collapse1, Allen2003collapse, Girart2009magnetic, Qiu2014submillimeter, Maury2018magnetically, Cortes2021magnetic, Bino2021fitting, Huang2024onthemagnetic, Saha2024magnetic}). On the other hand, weakly magnetized clouds typically have more chaotic magnetic fields (e.g., \citealp{Rao1998high, Beuther2010magnetic, Koch2018polarization, Tang2009evolution, Stone1998dissipation, Padoan2001turbulent, Zhang2014magnetic, Li2015magnetized, Hull2017alma, Koch2018polarization, Eswaraiah2021revealing, Wu2024ataleofthree}). However, the results of these observations are still insufficient to resolve the ongoing 
debate whether the key factor in the core formation, evolution, and its subsequent collapse to form a star is driven by a single key agent, magnetic field, or turbulence, or both? If both are important, what is their relative importance with respect to gravity? Furthermore, the observations capable of allowing a detailed analysis of the consistency of the role played by magnetic fields across multiple spatial scales remain scarce (e.g., \citealp{Li2009anchoring, Hull2014tadpol, Zhang2014magnetic, Li2015self, Hull2019interferometric, Liu2023ApJmultiscale}). 

Here, we present 1.3mm polarization observations with the Atacama Large Millimeter/submillimeter Array (ALMA) toward a high-mass star-forming region IRAS18360-0537. This region is located at a kinematic distance of 6.3 kpc \citep{Xu2021Water} with a far-IR luminosity of $1.2 \times 10^5$ L$_\odot$ \citep{Molinari1996asearch}. Two gas and dust condensations, MM1 and MM2, have been identified from the Submillimeter Array (SMA) 1.3 mm dust emissions. MM1 is revealed as a hot molecular core hosting an infalling envelope and an embedded O-type protostar with a current mass of $\sim 10 M_\odot$. A bipolar outflow is detected in CO and SiO emissions in a northeast-southwest (NE-SW) direction, which is approximately parallel with the rotation axis of MM1 (\citealp{Qiu2012forming}). \citet{Wu2023ALMA} presents an alternative interpretation, suggesting that a nearly east-west collimated jet is enclosed by an outflow cavity with an opening angle of approximately 140$^\circ$ as observed through ALMA CO and SiO emissions, as well as VLBA water maser spots. Here, we aim to reveal the magnetic field structure at high angular resolutions and investigate the importance of the magnetic field, gravity, and turbulence in the formation of high-mass stars in IRAS18360-0537. 

This study is part of the ALMA project aimed at imaging magnetic fields on scales of $10^3$ au envelopes. The observational results for the full sample and the analysis of the overall magnetic field properties are presented in the overview paper by \citet{Zhang2025star}. Furthermore, detailed studies of individual sources from this project have been partly published: the magnetic fields in G28.34+0.06 and NGC 6334 are presented in \citet{Liu2020magnetic, Liu2023ApJmultiscale}, respectively.

In Section 2, we describe the ALMA observations and data processing. In Section 3, we present observational results, showing the magnetic field morphology and providing quantitative analyses. We discuss the very intriguing magnetic field structure and provide possible explanations for the field structure in Section 4. Finally, we summarize the main conclusions in Section 5.\par

\section{OBSERVATIONS AND DATA ANALYSIS} \label{sec:obs}

\begin{table}[htbp]
    \caption{Observational Parameters} 
    \label{table:obsparameter}
    \centering
    \begin{threeparttable}     
	\begin{tabular}{ccccc}
		\hline\hline
		Date &  Configuration &  $N_{ant}$ & Min BL (m) & Max BL (m) \\ 
		\hline 		
		2018 Jun 23 & C43-1 &  47 & 15.0 & 312.7  \\ 
	    \hline 		
		2018 Jun 23 & C43-1 &  47 & 15.0 & 312.7  \\
	    \hline 		
		2018 Jun 29 & C43-1 &  47 & 15.1 & 313.7  \\
	    \hline 		
		2018 Sep 8  & C43-4 &  43 & 14.9 & 783.1  \\
	    \hline 		
		2018 Sep 9  & C43-5 &  47 & 15.0 & 1231.4 \\
	    \hline 		
		2018 Sep 12 & C43-5 &  43 & 15.0 & 1231.4 \\
		\hline 		
		2018 Sep 13 & C43-5 &  43 & 15.0 & 1231.4 \\
        
		\hline     	            
	\end{tabular} 
    
    \end{threeparttable} 
\end{table}

The full polarized ALMA observations of IRAS 18360-0537 (Project ID: 2017.1.00793.S, PI: Zhang) were conducted in 2018 June with the C43-1 configuration, and in 2018 September with the C43-4 and C43-5 configurations, across two observation sessions. Table \ref{table:obsparameter} lists the detailed information of the observations. The total on-source integration time is $\sim$20min. Band 6 receivers were employed for the project. The three Time Division Mode (TDM) basebands, each with a bandwidth of 1.875 GHz, are allocated to obtain the continuum emission at the central frequencies of 216.410, 218.516, and 233.517 GHz. The molecular line SiO(5-4) at a rest frequency of 217.105 GHz is covered in one of the continuum spectral windows with a channel width of 0.98 MHz. Four Frequency Division Mode (FDM) basebands were targeted at the molecular lines CO(2-1), OCS(19-18), $^{13}$CS(5-4), N2D$^+$(3-2). These spectral windows have a channel width of 122 kHz and a bandwidth of 58.6 MHz. However, all of these molecular lines failed to be detected due to an incorrect configuration of the systematic cloud velocity; therefore, the analysis of these lines is not included in this paper. At this frequency, the primary beam of the 12 m antenna was approximately $27^{\prime\prime}$. The flux and bandpass calibrations were performed on J1751+0939, and the phase calibrations were performed with J1851+0035. The quasar J1924–2914 was used for the calibration of instrumental polarization. The ALMA flux calibration accuracy in Band 6 is approximately $10\%$. \par

The datasets were calibrated and imaged using the Common Astronomy Software Applications (CASA; \citealp{McMullin2007casa}). We performed two iterations in phase-only and one iteration in amplitude and phase to self-calibrate on the Stokes I data. Self-calibration solutions were then applied to the spectral cubes. The imaging of Stokes I, Q, and U maps of the dust continuum was performed with the CASA task \textit{tclean}, using the Clark-Stokes deconvolution algorithm and the Briggs weighting with a robust parameter of 0.5. For each map, we applied the primary beam correction. The maps were imaged independently for the two observation sessions. For the data of C43-1 configuration session, the continuum maps yielded a synthesized beam of approximately 1$^{\prime\prime}$.37 $\times$1$^{\prime\prime}$.03 (PA. 50$^\circ$).  The 1 rms noises of the Stokes I maps reach 1.3 mJy beam$^{-1}$, while the Stokes Q/U maps have rms noises of 90 $\mu$Jy beam$^{-1}$. For the data of the C43-4 and C43-5 configuration session, the continuum maps have an angular resolution of approximately 0$^{\prime\prime}$.38 $\times$0$^{\prime\prime}$.30 (PA. 62$^\circ$) and shows an 1$\sigma$ rms of 0.35 mJy beam$^{-1}$ for Stokes I and 95 $\mu$Jy beam$^{-1}$ for Stokes Q and U maps.

The polarized images were computed from the quadrature sum of the Stokes Q and U intensity maps ($P_{biased}=\sqrt{Q^2+U^2}$). However, the polarized intensity $P$ always remains positive, even though the Stokes Q and U maps may have negative values. As a result, the measured values are biased towards larger values. To address this issue, we followed the debiasing method proposed by \citet{Vaillancourt2006placing}. The debiased polarized intensity $P$ can be obtained using the following equation:\par  
\begin{equation}
	P=\sqrt{Q^2+U^2-\sigma_{Q/U}^2}
\end{equation}

where $\sigma_{Q/U}$ is the average noise level of Stokes Q and U. For simplicity, we assume $\sigma_{P}\approx\sigma_{Q}\approx\sigma_{U}$ in the following calculations. Since this debiasing method provides better estimations for high signal-to-noise measurements \citep{Vaillancourt2006placing}, we limited our analysis to a threshold of $3\sigma_{P}$. The debiased polarization percentage $p$ and the polarization position angle $\theta$ can be calculated using:
\begin{equation}
	p=\frac{P}{I}
\end{equation}
\begin{equation}
	\theta=\frac{1}{2}tan^{-1}(\frac{U}{Q})
\end{equation}

Along with the continuum emission, we also detected molecular line transitions tracing different physical conditions. Those spectral lines were imaged using the CASA task \textit{tclean} with the \textit{Automasking} program \citep{Kepley2020auto-multithresh} and a Briggs parameter of 0.5. The data include the following transitions: SiO(5-4), CH$_3$OH(4$_2$-3$_1$), CH$_3$OH(5$_1$-4$_2$), CH$_3$OH(11$_3$-10$_2$), CH$_3$OH(18$_3$-17$_4$), CH$_3$OH(20$_1$-20$_0$), CH$_3$OH(10$_{2,9}$-9$_{3,9}$).

This work is mainly based on the analysis of the lower spatial resolution data (the C43-1 configuration session). For the higher spatial resolution (the C43-4 and C43-5 configuration session) data, our analysis is limited to the smaller-scale morphology of the magnetic field. Therefore, the continuum and molecular line maps hereafter represent those derived from lower resolution data, unless otherwise specified.

\section{Results} \label{sec:results}
\subsection{The Dust Continuum Emission and the Magnetic Field Morphology}
Figure \ref{fig:pibmap}(a) shows the total (Stokes I) and polarized dust emissions detected with the C43-1 observations ($\sim 1.2^{\prime\prime}$). The total emission reveals a filamentary structure oriented nearly north to south, and is dominated by a central bright source elongated along a northwest-southeast (NW-SE) direction. This bright source has been resolved into two dust condensations, MM1 and MM2, in previous sub-arcsecond ($\sim 1.5^{\prime\prime}$) SMA observations \citep{Qiu2012forming}, and also in our ALMA C43-4/5 map ($\sim 0.35^{\prime\prime}$, see Figure \ref{fig:bfieldmap}). In Figure \ref{fig:pibmap}(a), the polarized emission with signal-to-noise ratios (S/N) greater than 3 is detected toward the central and some northern parts of the filamentary structure. The polarization percentage spans from approximately 0.5\% to 6\%, with a median fraction of 3\%. Figure \ref{fig:pibmap}(b) shows the magnetic field orientations derived by rotating the polarization angles by 90$^{\circ}$, revealing a clear hourglass-shaped magnetic field around the central bright source, as well as the field with a northeast-southwest orientation in the north part of the filament. The average orientation of the hourglass field is aligned with the elongation of the central source, and is consistent with previous SMA polarization observations with a lower angular resolution ($1.7^{\prime\prime} \times 1.6^{\prime\prime}$) and lower sensitivity (1 $\sigma$ rms of 2.4 mJy beam$^{-1}$, see Figure \ref{fig:pibmap}(c)), but the hourglass-shaped structure is only seen in our ALMA data. 

Figure \ref{fig:bfieldmap} shows the total dust emission and magnetic field orientation maps obtained with the ALMA C43-4/5 observations ($\sim$ 2000 AU). At a factor of 3.5 times higher angular resolution ($\sim$ 10000 AU), the polarized emission is only detected toward the two dust condensations, MM1 and MM2, and shows a more uniform magnetic field with its orientation well aligned with that of the inner part of the hourglass field. On a larger scale ($\sim$0.5 pc), the magnetic field permeating the filamentary gas structure also has an NW-SE orientation (private communication with the BISTRO consortium). Therefore, the magnetic field has a mean orientation all the way aligned from the 0.01 pc (revealed by the ALMA C43-4/5 data) to 0.04 pc (revealed by the ALMA C43-1 data and the SMA data) and 0.5 pc scales.

\subsection{The Molecular Line Emission}
 In addition to the continuum emission, we detect several molecular spectral lines in emission with the 1.875 GHz bandwidth. Figure \ref{fig:sioandch3ohmap} shows the SiO (5-4) and CH$_3$OH ($10_{2,9}-9_{3,6}$) emissions; the former traces outflow shocks, and the latter probes the kinematics of the dense gas. A northeast-southwest (NE-SW) bipolar outflow centered at MM1 is seen in the SiO map, and is consistent with the previous SMA observations \citep{Qiu2012forming}. The CH$_3$OH emission reveals a clear NW-SE velocity gradient around MM1 and MM2, and the gradient is nearly perpendicular to the outflow axis. We note that other CH$_3$OH lines are also detected in the bandwidth and reveal similar kinematics. We refer the readers to \citet{Mo2023exploring} for a more detailed analysis and interpretation of the CH$_3$OH data.

\section{Discussions} \label{sec:discussions}
\subsection{Does hourglass shape morphology represent a strong field ?}
The detection of an hourglass-shaped magnetic field is often interpreted as evidence of core collapse regulated by a strong magnetic field \citep{Allen2003collapse, Girart2006magnetic, Girart2009magnetic, Hull2019interferometric}. This interpretation, however, implies an underlying assumption of alignment between the mean magnetic field orientation, the pseudo-disk rotational axis, and the outflow axis, based on the classic magnetically regulated collapse model. In this model, the flux-frozen magnetic field is dragged into an hourglass configuration by the collapsing gas, providing considerable support against gravity. During the collapse process, angular momentum is transported outward from the core via magnetic braking: the field lines connect the fast-rotating pseudo-disk with the more static envelope, and the braking torque exerted by magnetic tension transfers angular momentum from the pseudo-disk to the envelope \citep{Allen2003collapse}. In the presence of a strong magnetic field, this effect becomes more significant, causing most of the accreting material to flow inward through a pseudodisk, while aligning the disk rotation axis and the outflow with the magnetic field direction (\citealp{Allen2003collapse, Machida2006second, McKee2007theory, Girart2006magnetic, Sadavoy2018dust1623a, Cortes2021magnetic}). Therefore, as a dense core evolves further and a disk forms within it, driving an outflow, the outflow and rotation axes tend to align with the magnetic field orientation. However, the ideal magnetic braking scenario predicts small disks due to its highly braking efficient, and the consideration of various non-ideal MHD effects is a possible solution \citep{Dapp2012bridging, Li2014ontherole}.

\begin{figure}[htbp]
	\centering
    \begin{minipage}[b]{0.48\linewidth}
        \centering
        \includegraphics[width=1\linewidth]{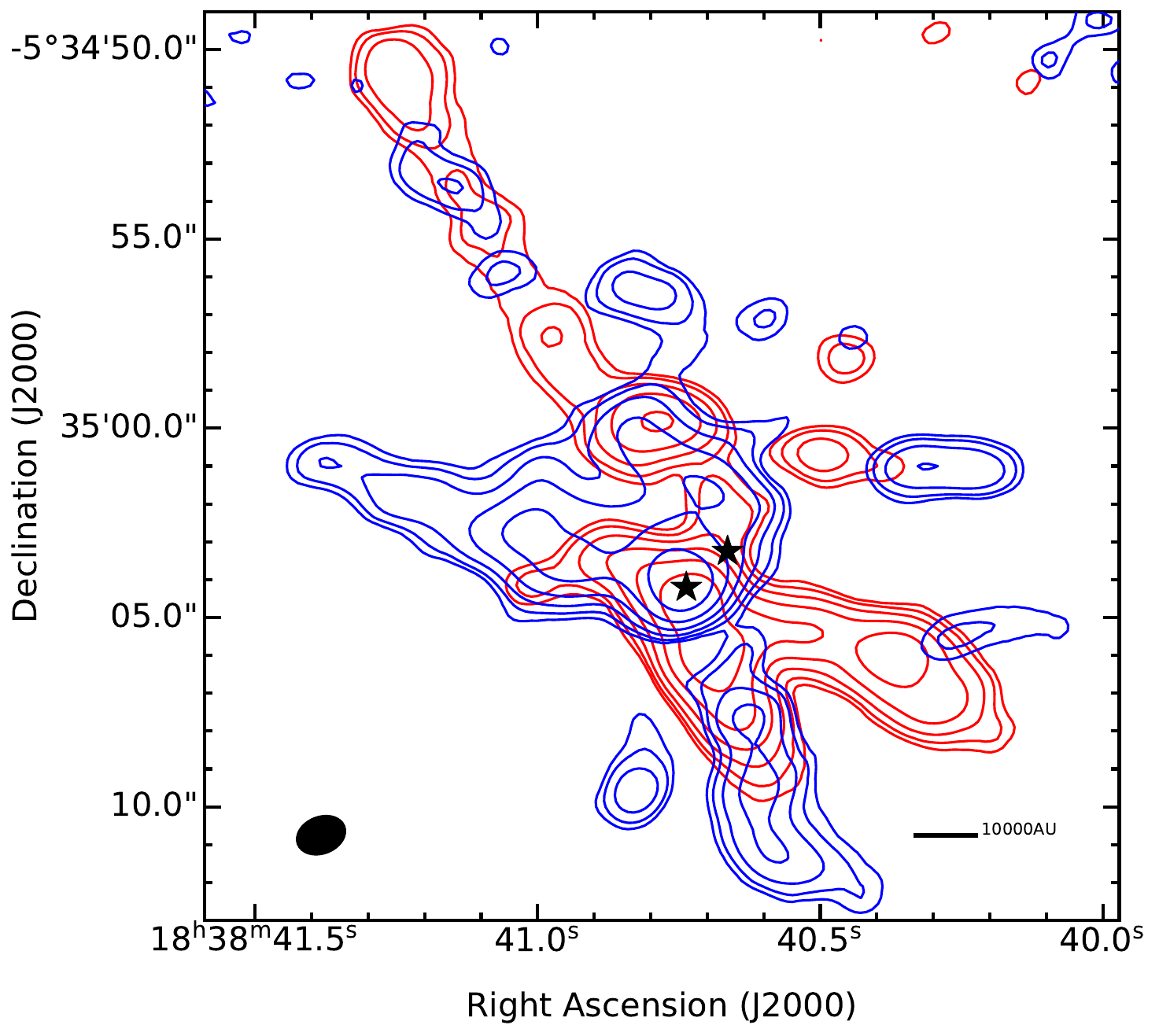}
    \end{minipage}
    \hfill
    \begin{minipage}[b]{0.48\linewidth}
        \centering
        \includegraphics[width=1\linewidth]{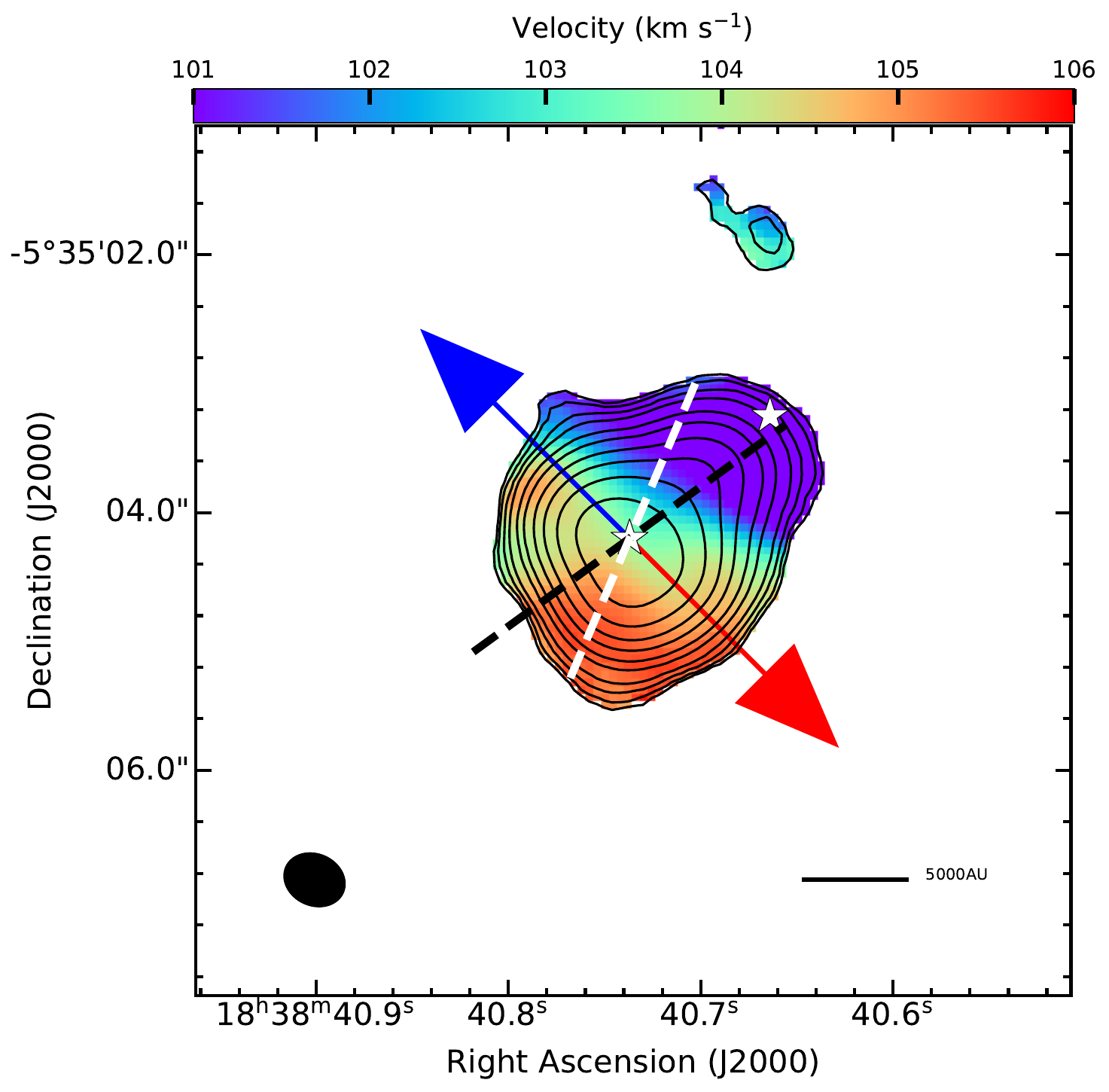}
    \end{minipage}
 
    \caption{(a) The SiO (J=5-4) emission observed with the ALMA and integrated from 85.8 to 100.6 kms$^{-1}$ for the blue-shifted lobe and from 106.0 to 120.8 kms$^{-1}$ for the red-shifted lobe, shown in blue and red contours, respectively. The contour levels of SiO emission are 3, 5, 9, 17, 33, 65, 129, 257 times $\sigma$ = 36 mJy beam$^{-1}$ km s$^{-1}$. (b) The color image shows the first-moment map of the CH$_3$OH ($10_{2,9}-9_{3,9}$) emission, and the contours correspond to the zeroth moment contour map. The contour levels are 3, 5, 9, 17, 33, 65, 129, 257, 513 times $\sigma$ = 8 mJy beam$^{-1}$ km s$^{-1}$. The black dotted line represents the velocity gradient of the CH$_3$OH map, and white dotted lines are the gradient identified in \citet{Wu2023ALMA}. The blue and red arrows represent the bipolar outflow axis.}
	\label{fig:sioandch3ohmap}
\end{figure}

\begin{figure}[htbp]
	\centering
	\includegraphics[width=0.6\linewidth]{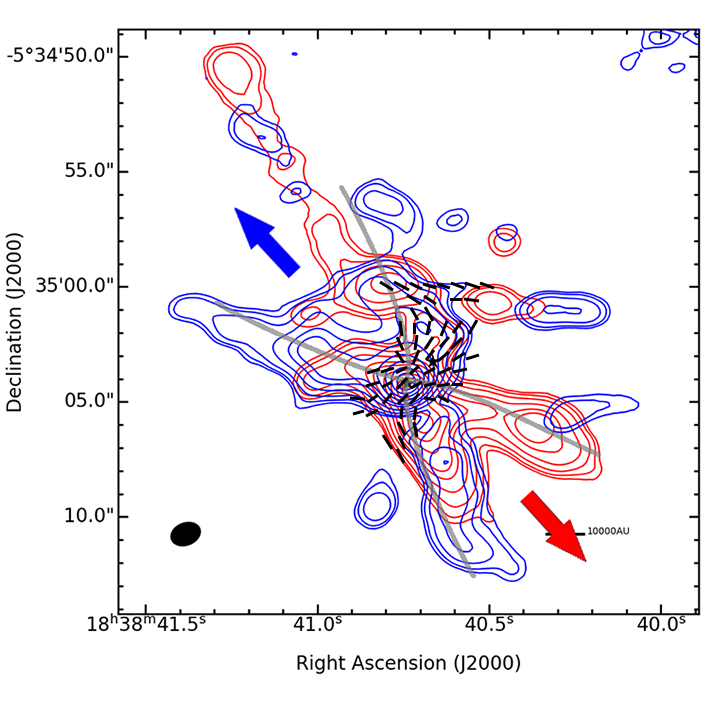}
    \includegraphics[width=0.4\linewidth]{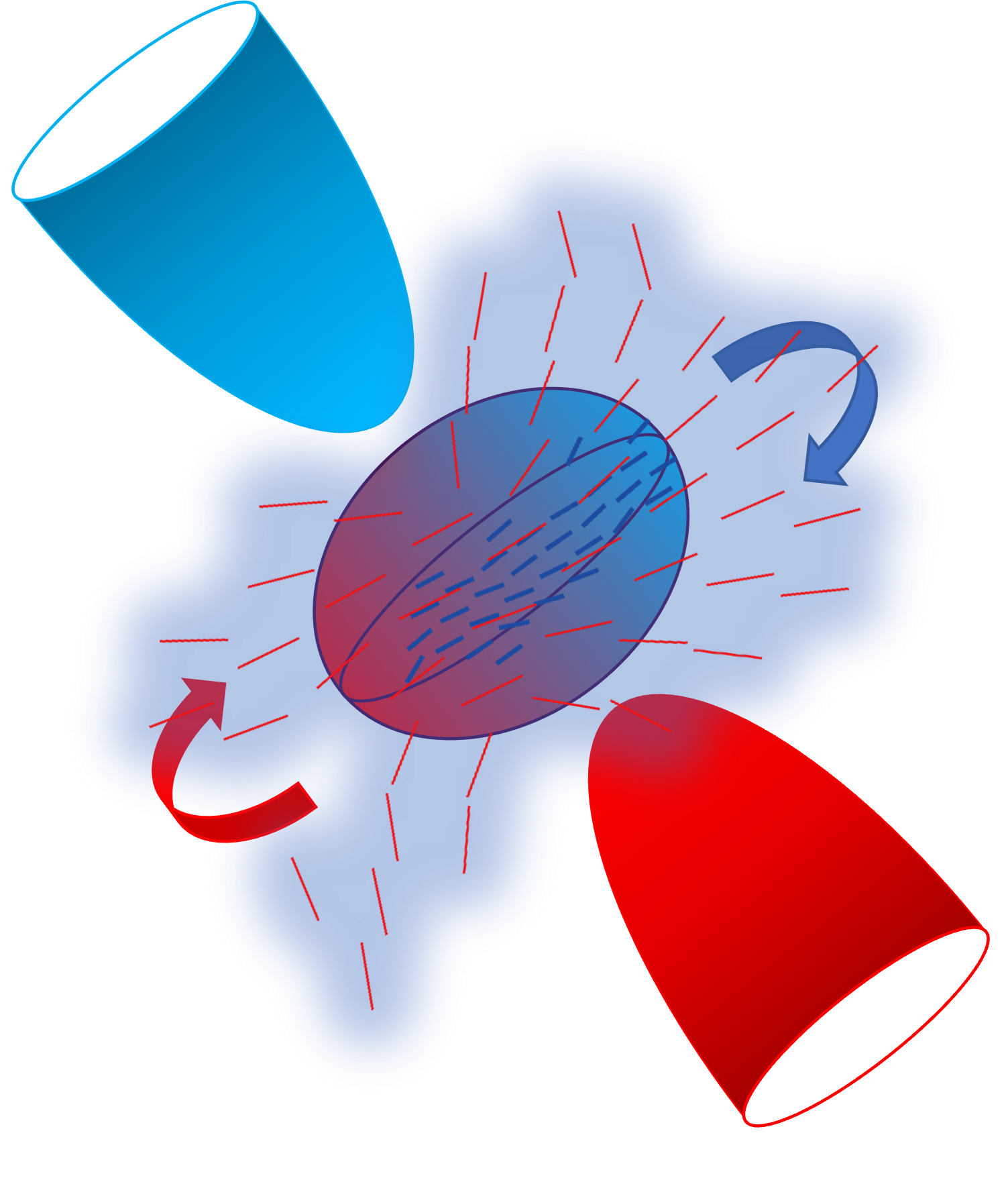}
	\caption{Upper Panel: The magnetic field lines (black segments) overplotted on the SiO (J=5-4) emissions (colored contours). The red and blue contours are the same as Figure \ref{fig:sioandch3ohmap}(a). The gray curves outline the bipolar outflow, and the red and blue arrows indicate the orientation of the outflow axis. The observed magnetic lines, situated away from the central region, are consistent with the cavity of outflows, suggesting that the magnetic field is influenced by outflows. Lower Panel: This schematic presents a potential explanation for the morphology of the magnetic field. The envelope scale field lines are dragged into a spiral configuration by the strong gas rotations.}
	\label{fig: magnetic_outflow}
\end{figure}

Figure \ref{fig:sioandch3ohmap} shows bipolar outflow and the velocity maps of its corresponding dense gas structures. The outflow and rotation axes are well aligned, also consistent with previous molecular observations \citep{Qiu2012forming, Wu2023ALMA}. However, the most intriguing result in this work is that such a highly “ordered” hourglass-shaped magnetic field is almost perpendicular to the outflow/rotation axis. The upper panel of Figure \ref{fig: magnetic_outflow} shows the magnetic orientations overlaid on the SiO (5-4) outflow. This misalignment apparently contradicts the predictions of the classic magnetically regulated collapse model at the scales probed by our ALMA and SMA observations, i.e., the scales ranging from several thousand to tens of thousands of AU. On the other hand, the hourglass-shaped magnetic field structure is very ordered, and thus inconsistent with the predictions of weak-field models. Of course, conversely, if we insist on adopting such a classical strongly magnetized collapse model, we might assume that at smaller scales, the disk scale magnetic field on scales of a few hundred AU, where the outflow originates, is aligned with the outflow direction. This implies that the condenstaion scale magnetic field at our current resolution limit ($\sim2000$AU) is decoupled from the disk-scale (several hundred AU) magnetic field (\citealp{Chandler2005stars, Hull2017alma, Myers2020magnetic, Pattle2021omc}). Nevertheless, this predictive interpretation requires higher-resolution observations and is beyond the scope of this paper. In any case, at least on condensation scales, we cannot explain the perpendicularity between the magnetic field and the outflow using the classical model.

Gravity and gas bulk motions are key factors influencing the magnetic field morphology at multiple physical scales. Our discussion proceeds from large to small scales. At the filament scale, numerous studies of magnetic field morphology focus on comparing the magnetic field direction with the filament orientation. Observations ranging from sub-millimetre dust continuum polarization to optical/NIR starlight polarization show that lower-density filaments or striations tend to align parallel to the magnetic field, while dense filaments preferentially align perpendicular to the local magnetic field direction \citep{Goldsmith2008large, Chapman2011magnetic, Palmeirim2013herschel, Franco2015tracing, Panopoulou2016magnetic, PlanckCollaboration2016planck, Santos2016magnetically, Kusune2019magnetic, Sugitani2019nearinfrared, Arzoumanian2021dust}. Using the Histogram of Relative Orientations (HRO) method, \citet{PlanckCollaboration2016planckintermediate} statistically quantified this change in alignment across 10 nearby molecular clouds and found that the transition from parallel to perpendicular alignment occurs at a critical column density of $log_{10}(N_H/cm^{-2})\approx 21.7$. A possible interpretation of this transition is that in low-density regions the magnetic field is relatively stronger and imposes a preferred gas flow direction parallel to the field, whereas as the density increases and gravity takes over, the magnetic field provides support against isotropic gravitational collapse \citep{Soler2017whatare, Seifried2020fromparallel}. However, some observations and simulations point toward a second transition: when filaments or dense cores become highly supercritical, the preferentially perpendicular magnetic field orientation returns to parallel as the field lines are dragged by the accreting gas along the filament \citep{Mocz2017movingmesh, Liu2018compressed, Pillai2020magnetized, Kwon2022bfields}. At core scales of $\lesssim 0.1$ pc, the morphology of the magnetic field exhibits greater complexity. Numerous observations reveal very ordered magnetic field structures, indicating a dominant role played by magnetic fields \citep{Girart2006magnetic, Stephens2013magnetic, Qiu2014submillimeter, Hull2014tadpol}. However, disordered to chaotic magnetic field structures, which suggest an overwhelming role played by turbulence, have also been observed (e.g., \citealp{Zhang2014magnetic}). Spiral magnetic field patterns have also been observed, where gravity and rotation play a significant role in shaping the magnetic field \citep{Lee2019pseudodisk, Beuther2020gravity, Sanhueza2021gravity}. Most notably, G192.16-3.84 \citep{Liu2013gas} and NGC 6334 V \citep{Juarez2017magnetized} exhibit magnetic field morphologies similar to our source, where the field direction is roughly parallel to the gas velocity gradient. For G192.16-3.84, the authors propose that gas rotation or infall influences the magnetic field direction, while for NGC 6334 V, they argue, based on comparison with MHD simulations, that gravity-dominated gas infall drags the magnetic field lines. When it comes to disk scales of $\lesssim 1000$ AU, magnetic field morphologies are more strongly affected by rotation, accretion flows, and outflows. Moreover, at these scales, polarized emission is no longer dominated by dust grains aligned via the radiative alignment torques (RAT) mechanism \citep{Lazarian2007radiative}. Instead, it may be contaminated by self-scattering from large dust grains (Kataoka et al.), rendering the magnetic field morphology at disk scales difficult to detect. \citet{Sadavoy2018dust1623a} analyzed the polarization maps toward VLA 1623A and suggested that the polarization in the central disk region arises from dust self-scattering, while that in the surrounding ring-like structure remains from RAT-aligned dust grains. Therefore, the magnetic field of VLA 1623A reveals an hourglass shape on scales of approximately 200 AU. In contrast, L1448 IRS2 and BHB07-11 exhibit distorted hourglass structures; the distortion observed in L1448 IRS2 arises from dragging by accretion flows or outflows \citep{Kwon2019highly}, whereas BHB07-11 is best modeled with a dominant poloidal component and a weaker toroidal component \citep{Alves2018magnetic}. The magnetic field geometry observed in B335 can be explained by a family of MHD models, which suggest that initially planar field lines are dragged along the direction of gravitational collapse \citep{Kandori2020distortion}. In short, in dense molecular clumps and cores, protostellar envelopes and disks, gravitational infall, accretion, rotation, and outflows could significantly alter and twist the magnetic field morphologies.

\begin{figure}[htbp]
	\centering
	\includegraphics[width=1\linewidth]{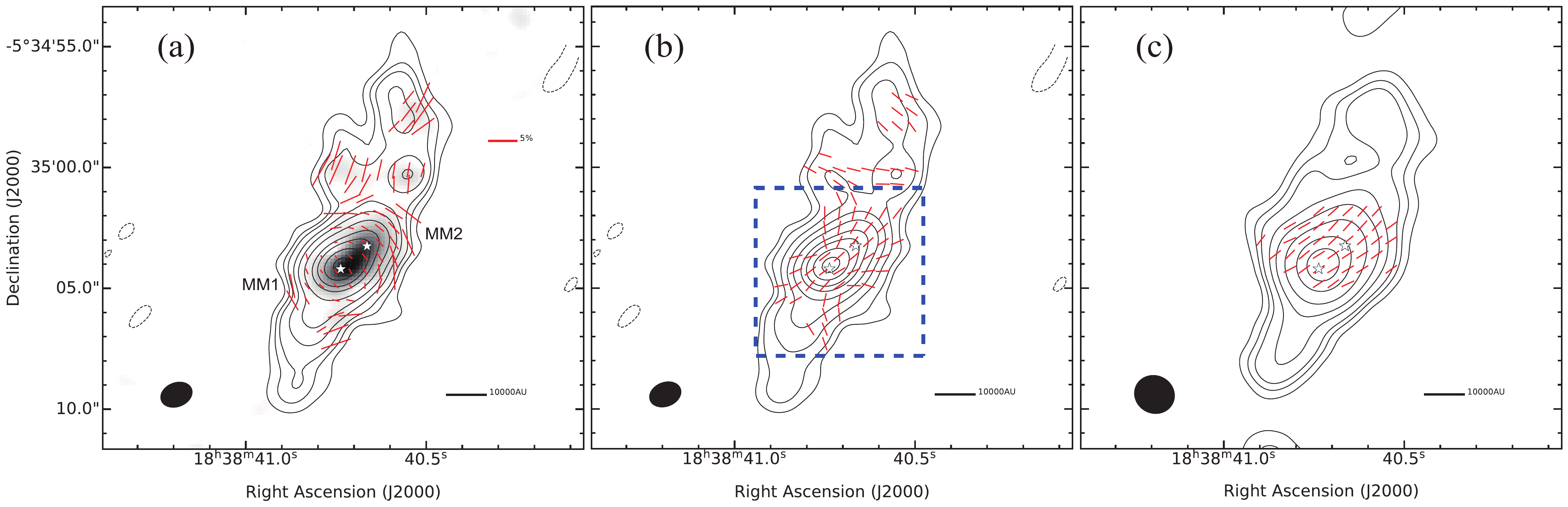}
	\caption{(a) ALMA 1.3 mm dust continuum contours toward IRAS18360 overlaid polarization intensity maps (grayscale) and polarization orientations at lower resolution ($\sim$1.2$^{\prime\prime}$). The contour levels are -2, 3, 5, 7, 12, 20, 40, 90, 150, 260, 410 times the $\sigma$ value of 1.3 mJy beam$^{-1}$. The polarization segments (drawn following the Nyquist sampling) are plotted above the 3$\sigma$ level ($\sigma$ = 95 $\mu$Jy beam$^{-1}$), with segment lengths proportional to polarization fraction (5\% reference at top-right). Stars mark the position of dust condensations MM1 and MM2 \citep{Qiu2012forming}. (b) Contours and stars as in (a), but segments show magnetic field orientation (rotating the polarization segments by 90$^\circ$). The blue rectangle represents the same region shown in Figure \ref{fig:bfieldmap}. (c) The 1.3 dust continuum contours and magnetic field segments observed by 1.3mm SMA polarization observations \citep{Zhang2014magnetic}. The contour levels are 3, 5, 7, 12, 20, 40, 90, 150 $\times$ $\sigma$ = 7.5 mJy beam$^{-1}$. The synthesized beam size and scale bar are shown at the bottom of each panel.}
	\label{fig:pibmap}
\end{figure}

We therefore propose that the magnetic fields in IRAS 18360 are strongly influenced by the gravity and bulk motions of the gas, particularly outflows, rotations, and infalls, as illustrated in the lower panel of Figure \ref{fig: magnetic_outflow}. A key indirect piece of evidence is the pronounced radial gradient in polarization percentage shown in Figure \ref{fig:pibmap}. This gradient could indicate the presence of beam smearing or a line-of-sight averaging effect for the orientations near the central region. This suggests that the magnetic field is likely less ordered on smaller spatial scales, or along the line of sight, than the smoothed configuration we currently observe. Second, we see that the magnetic field appears to be correlated with the outflow cavity wall to the southwest and to the north of the outflow center, similar to what is observed in other interferometric studies \citep{Ching2016helical, Hull2017alma, Maury2018magnetically, Hull2019interferometric, LeGouellec2019haracterizing, Encalada2024magnetic}. This is consistent with the comparison between the outflow energy and the magnetic energy presented in Section \ref{Qcompartion}. A proposed explanation for such detection is that outflows compress gas along the cavity walls, increasing its density beyond the critical value required for polarization detection \citep{Ching2016helical}.

\begin{figure}[htbp]
	\centering    
    \includegraphics[width=1\linewidth]{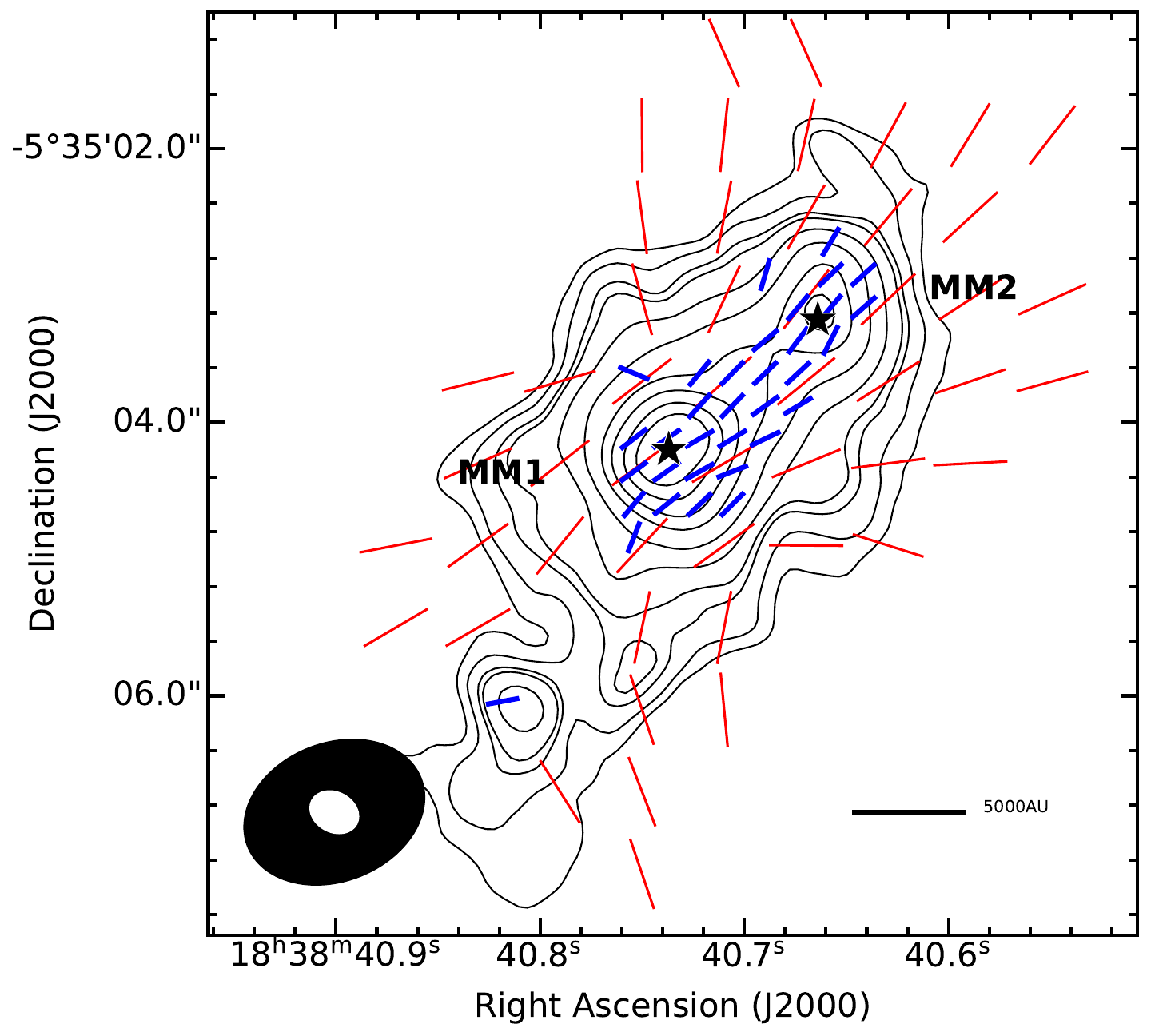}
    
    \caption{Zoomed-in view of the blue rectangle area in Figure \ref{fig:pibmap}. Contours show higher-resolution ($\sim$0.35$^{\prime\prime}$) dust continuum emission in steps of 5, 7, 12, 20, 40, 90, 150, 260, and 410 times the $\sigma$ value of 350 $\mu$Jy beam$^{-1}$. Blue segments represent the magnetic field orientation observed in higher resolution. Red magnetic segments are replicated from Figure \ref{fig:pibmap}(b).}
	\label{fig:bfieldmap}
\end{figure}

Additionally, the central nearly uniform magnetic field structure is well aligned along the axis connecting the two condensations, with this alignment becoming more pronounced in higher-resolution ALMA observations (Figure \ref{fig:bfieldmap}). We propose that this alignment is caused by gravitational pull from higher-density gas drags and reorients the magnetic field. The polarization–intensity gradient technique analysis \citep{Koch2012magnetic} indicates that gravity dominates the magnetic field (see Appendix \ref{app:compBGT}). Consistent with this, alignment analysis of this ALMA project sample shows that gravity in high-density regions tends to drag magnetic fields, resulting in alignment between the field and the direction of gravitational forces \citep{Zhang2025star}, although IRAS 18360 is excluded from this analysis due to its large distance. Among gravity-driven processes, gas accretion is a possible mechanism for influencing the magnetic field geometry. Given that the fainter condensation, MM2, likely represents a fragment of the envelope \citep{Qiu2012forming} and may be exchanging material with the central protostar. Finally, attributing the core’s velocity gradient solely to rotation would imply significantly higher rotational energy than typical values \citep{Goodman1993angular}. This suggests that the rotating gas could pull the field lines into a toroidal configuration. However, we do not exclude the possibility of a combined effect from both infalling and rotating motions. In the following section, we further investigate this interpretation by presenting quantitative analyses.

\subsection{Quantitative Comparisons}\label{Qcompartion}
To quantitatively compare the relative importance of the outflow and rotation with the magnetic field, we estimate the magnetic field strength (see Appendix \ref{app:dcf}) and use its upper limit ($B_{pos,u} = 1.7$ mG) in our comparison. 

We first estimated the energetics of the outflow. Because our dataset lacks CO line data, we could not directly derive the outflow mass and kinetic energy. Instead, we used the SiO emission to trace its dynamical properties, since the SiO morphology is consistent with that presented in \citet{Qiu2012forming}. Given the poorly constrained SiO abundance, we adopted two independent approaches to estimate the outflow mass. First, we adopted the outflow mass of $\sim 54 M_\odot$ obtained by \citet{Qiu2012forming} from combined SMA and IRAM 30m CO data. Second, we applied the empirical relation between outflow mass and clump mass $M_{out} =0.3 M_{clump}^{0.8}$ \citep{Beuther2002massive, Lopez-Sepulcre2010comparative}, which yields an outflow mass of about $10 M_\odot$. From these two mass estimates, we computed the corresponding kinetic energies to be $4.3 \times 10^{46}$ erg and $8.0 \times 10^{45}$ erg, respectively. Both values exceed the magnetic energy of $4.6 \times 10^{45}$ erg derived from the $1.7$ mG upper limit of magnetic field strength.

From a dimensional point of view, the centrifugal force and the radial component of the Maxwell stress tensor are comparable, which implies that the rotational energy is comparable to the magnetic energy. We therefore proceed to compare the magnetic and rotational energies. Assuming spherical moment of inertia, the average ratio of rotational to magnetic energy can be expressed as: $\beta_{r-B}=2/3\cdot(3-n)/(5-n)(v_{rot}/V_A)^2$ \citep{Sanhueza2021gravity}. Here, $v_{\text{rot}} = \Omega \cdot R$ is the rotational velocity under the assumption of solid-body rotation, where $\Omega$ and $R$ are the angular velocity and radius, respectively. The parameter $n$ is the power-law index of the density profile, defined as $\rho \propto R^{-n}$. We use the high-excitation CH$_3$OH(10$_{2,9}$-9$_{3,9}$) line to get the velocity field. This line reveals a clear SE-NW velocity gradient aligned with the cloud's elongation (Figure \ref{fig:sioandch3ohmap}). Assuming this gradient purely represents rotation, we estimate the average rotational velocity to be 2.4 km s$^{-1}$. The Alfv\'{e}n velocity derived from the magnetic field strength is $V_A \sim 2.5 $km s$^{-1}$. We obtain a $\beta_{r-B}=E_{rot}/E_B$ ratio of 0.44 for a uniform density distribution($n=0$) and 0.29 for a density profile with $n$=1.5. 

\citet{Machida2005collapse} proposed that the ratio of angular velocity to magnetic field strength $\omega/B$ can serve as a criterion to evaluate whether centrifugal or magnetic forces govern the evolution of a collapsing core. An observed ratio exceeding the critical value $(\omega/B)_{crit}=0.39\sqrt{G}/\sigma_{th}$ indicates rotational dominance over the magnetic field; conversely, the magnetic field dominates if the ratio is lower. For dust temperatures of 55 K and 160 K (as discussed in Appendix \ref{app:temp_discussion}), the critical values are $\sim 7.1 \times 10^{-8} \text{ yr}^{-1} \mu\text{G}^{-1}$ and $\sim 4.2 \times 10^{-8} \text{ yr}^{-1} \mu\text{G}^{-1}$, respectively. Our evaluation of the observed ratio gives $(\omega/B)_{obs} \sim 2.0 \times 10^{-8} yr^{-1} \mu G^{-1}$.

Both methods suggest the magnetic field is marginally more significant than rotation.  However, it is crucial to note that the magnetic field strength estimated via the Davis-Chandrasekhar-Fermi (DCF) method, as shown in the appendix \ref{app:dcf}, is likely significantly overestimated. Consequently, rotational energy probably plays a more dominant role than these initial comparisons suggest. This apparent discrepancy arises partly because the rotational velocity is averaged over the entire magnetic field sampling region. Notably, within the smaller-scale outflow-driving core MM1, rotation is faster and dynamically more significant.

An alternative approach to assessing rotational significance is to evaluate the rotation-to-gravity energy ratio, $\beta_{r-G}=(5-2n)/(5-n)({v_{rot}^2 R}/{3GM})$ \citep{Sanhueza2021gravity}. This ratio avoids complications from magnetic field measurements. Observations show cores typically rotate slowly, with $\beta_{r-G} \lesssim 0.15$ (\citealp{Goodman1993angular}). Multiscale studies report a median $\beta_{r-G} \sim 0.03$ (\citealp{Goodman1993angular, Caselli2002dense, Belloche2013complex, Chen2019droplets}). For IRAS 18360, $\beta_{r-G}$ values are 0.34 (uniform density profile $n$=0) and 0.19 ($n$=1.5), exceeding the median value by nearly an order of magnitude. This indicates that rotation motions are significant in this region.

\begin{figure}[htbp]
	\centering
	\includegraphics[width=0.8\linewidth]{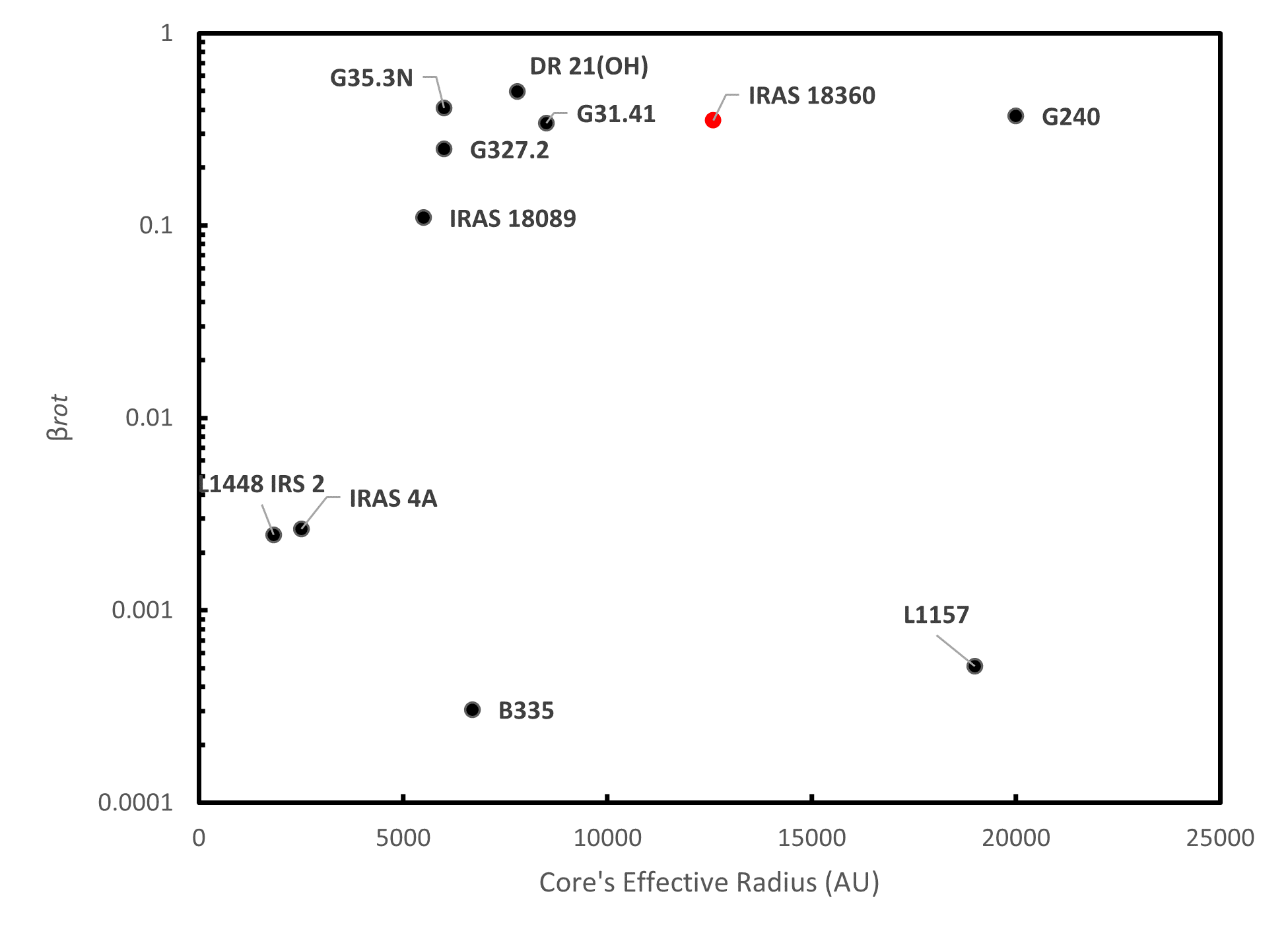}
	\caption{Ratio of rotational energy to gravitational energy $\beta_{rot}$ vs. core radius. Data for the sources shown are taken from \citet{Hull2019interferometric} and \citet{Pattle2021omc} except IRAS 18360 (this work). We have investigated those magnetic-detected sources as samples and ultimately selected those with the detection of velocity gradients or rotations, which are: L1448 IRS2 \citep{Kwon2019highly}, NGC1333 IRAS4A \citep{Attard2009magnetic}, B335 \citep{Yen2019JCMT}, L1157 \citep{Tobin2011complex, Sharma2020distance}, IRAS 18089 \citep{Sanhueza2021gravity}, G327.3 \citep{Beuther2020gravity}, G35.2N \citep{Qiu2013from}, DR 21(OH) \citep{Girart2013dr21}, G31.41 \citep{Girart2009magnetic, Beltran2019ALMA}, G240.31 \citep{Qiu2014submillimeter}. We note that the values of $\beta_{rot}$ in this plot are lower limits, as they are not corrected for inclination.}
	\label{fig: rotation_gravity_ratio}
\end{figure}

In Figure \ref{fig: rotation_gravity_ratio}, we revisit the dense cores with well-characterized magnetic field morphologies and strengths (same as the source in \citep{Hull2019interferometric, Pattle2023magnetic}. The resulting median $\beta_{r-G}$ for these cores is 0.04, consistent with the aforementioned statistical results. IRAS 18360 exhibits a closer affinity in its $\beta_{r-G}$ ratio to cores where rotation significantly impacts the magnetic field than to magnetically dominated cores. Examples include: The nearly edge-on high-mass core IRAS 18089-1732 \citep{Sanhueza2021gravity}, which displays a spiral-like magnetic field morphology at scales of several thousand AU, has a ratio of 0.11. Observations of twisted magnetic field lines in the high-mass core G327.2 at 6000 AU scales \citep{Beuther2020gravity} correspond to a lower-limit ratio of 0.29, derived from comparing rotational energy within the twisted field radius to the gravitational energy of the entire region. DR21(OH) \citep{Girart2013dr21}, where a toroidal magnetic field yields a ratio of 0.5 at the corresponding scale (7800 AU), and the filamentary clump G35.2N with a rotationally distorted field has a ratio of 0.4 \citep{Qiu2013from}. Conversely, most dense cores exhibiting clear hourglass magnetic field morphologies, along with alignment between field, rotation, or outflow axes, show $\beta_{r-G}$ values near the median value. G240 is an exception, where $\beta_{r-G}$ may be overestimated due to condensations along the velocity gradient \citep{Qiu2014submillimeter}. Therefore, assuming $\beta_{r-G}$ effectively reflects rotational importance, the derived value of about 0.4 for IRAS 18360 suggests rotation can twist magnetic fields into toroidal morphologies.

Following the energy analysis indicating significant rotational motion in this source, we provide further evidence for such rotation by applying centroid fitting to the CH$_3$OH(10$_{2,9}$-9$_{3,9}$) line. The position-velocity curve is shown in Figure \ref{fig: centriod_fitting}. The positional accuracy of the centroid velocity exceeds the interferometer's resolution by approximately a factor of the signal-to-noise ratio. More precisely, the total uncertainty in the centroid positions also accounts for a contribution from the bandpass. This yields $\sigma_{centroid} = (\sigma_{fit}^2 + \sigma_{BP}^2)^{1/2}$, with $\sigma_{fit}=\theta_{beam}/(2 SNR)$ and $\sigma_{BP}=\theta_{beam}(\Delta\Phi_{BP}/360^{\circ})$ \citep{Zhang2017angular}. The
phase noise in the bandpass calibrator is $\Phi_{BP}\approx 3^{\circ}$ for the spectral window of this line. A nearly linear velocity gradient is observed between 98 km s$^{-1}$ and 110 km s$^{-1}$, consistent with a pseudo-disk-like structure.  The distance from the polar axis was calculated as the shortest distance to the best-fit axis. This axis passes through the continuum peak and is orthogonal to the velocity gradient defined by the centroid velocities. Following \citet{Zhang2017angular}, we assume that the transitions are localized on a fixed-radius disk, where the slope of the centroid velocity plot can be described as $v_{cen}=\sqrt{GM/R^3} l$. $G$ is the gravitational constant, $R$ is the disk radius and $l$ is the distance from the polar axis \citep{Prasad2023detection}. We adopt the pseudo-disk radius as the maximum distance of the centroid velocity, $R\sim 2500$ AU. However, as shown in Figure \ref{fig: centriod_fitting}, the velocity gradients on the blue-shifted and red-shifted lobes are different. We therefore performed separate linear regression on the blue-shifted and red-shifted position-velocity slopes, obtaining an estimated enclosed mass of 49.6$\pm$1.1$M_\odot$ to 162.4$\pm$7.3$M_\odot$, respectively. This discrepancy likely arises because the disk is not perfectly symmetric. The blue-shifted region might be affected by condensation MM2, making its effective radius larger than assumed. As a result, the blue-lobe mass is underestimated, and the red-lobe mass is overestimated. These estimates are consistent with the protostar mass of about 60$M_\odot$ fitted by our radiative transfer simulations \citep{Mo2023exploring}. This kinematic evidence confirms strong rotational motion in this source and supports our previous view that the magnetic field lines are significantly influenced by the gas rotations.

In addition to rotation, gas motions driven by gravity can also influence the magnetic field morphology. As shown by \citet{Qiu2012forming}, the core's kinematics includes both Keplerian-like rotation and infall motions. Radiative transfer simulations of the CH$_3$OH(10$_{2,9}$-9$_{3,9}$) line \citep{Mo2023exploring} further confirm that the core is both rotating and infalling. Based on the \citet{Mo2023exploring} model, rotational and infall energies constitute 57\% and 43\% of the total kinetic energy, respectively, implying that the infall of gas cannot be ignored. Furthermore, the magnetic field observed at a sub-arcsec resolution is well aligned with the bridge connecting the two condensations. Similar magnetic configurations occur in both low- and high-mass cores such as NGC6334V and IRAS16293 (\citealp{Juarez2017magnetized, Sadavoy2018dust16293}). Such field morphology can be explained by the influence of gas flows. Therefore, we propose an alternative explanation that gas accretion influences the magnetic field lines. Following \citet{Sadavoy2018dust16293}, we model accretion as a cylindrical flow from MM2 to MM1's protostar. The cylinder length, representing the distance between the two condensations' centers, is $2L=9500$ AU \citep{Qiu2012forming}. The gravitational acceleration on a test particle at a displacement $x$ from the center of the cylinder  is as follows:
\begin{equation}
    g=\dfrac{GM_1}{(L-x)^2} -\dfrac{GM_2}{(L+x)^2}
\end{equation}
where $M_1$ and $M_2$ denote protostellar masses of MM1 and MM2, respectively. As the particle falls toward the central star, it gains kinetic energy. The corresponding gas flow velocity is given by:
\begin{equation}
    v_{flow}^2  =v_0^2  + 2G M_1 \left[ \dfrac{1}{L-x}-\dfrac{1}{L-x_0} \right] + 2 GM_2\left[ \dfrac{1}{L+x}-\dfrac{1}{L+x_0} \right]
\end{equation}

Here, $v_0$ represents the initial velocity at position $x_0$. For simplicity, we derive flow velocities for two limiting cases. First, when the gravitational influence of MM2 is negligible, the flow velocity starting from rest at MM2 is given by $v_{flow}^2={2G M_1}/L [1/(1-x/L)  -1/2]$. Conversely, assuming that the protostar mass in MM2 equals to that in MM1, the flow velocity from rest at the midpoint becomes $v_{flow}^2 = 4G M_1 /L [1/(1-(x/L)^2 ) -1]$. The latter expression agrees with Equation 9 in \citet{Sadavoy2018dust16293}. Expressing the flow velocity as $v_{flow}=a \sqrt{GM_1 /L}$, where the constant $a$ depends on position, we find $a$ ranges from 0.6 at $x=-L/2$ to 1.7 at $x=L/2$ for the first case. For the latter case, $a$ varies from 0.5 to 2.3 as $x$ shifts from $L/4$ to $3L/4$, remaining within the same order of magnitude in both scenarios. Consequently, for a protostar mass $M_1=10M_\odot$ \citep{Qiu2012forming}, we estimate $v_{flow} \sim \sqrt{G M_1/L} \sim$ 1.4 km s$^{-1}$. Moreover, our modeling in \cite{Mo2023exploring} provides an alternative estimate of the infall velocity. The model assumes an envelope 80 $M_\odot$ with a density profile following $\rho\propto r^{-2.5}$. The infall velocity profile is given by $v_{infall}(r)=v_0\sqrt{r_0(r-r_0)}/r$, with inner envelope radius $r_0=730$ AU and outer radius $R=12600$ AU. Here, $v_0$ is the Keplerian velocity at $r_0$ for a 60 $M_\odot$ protostar. Weighting the velocity by $\sqrt{\rho}$ to reflect the total infall energy, we estimate the average infall velocity of $\bar{v}_{model}\simeq $1.9 km s$^{-1}$, in agreement with both above toy model and the 1.5 km s$^{-1}$ value derived from CN (2–1) line profile fitting \citep{Qiu2012forming}. The similarity between the accretion/falling gas velocity and the rotational velocity supports the idea that gas flows may play a significant role in shaping the magnetic field morphology.
 
\subsection{Magnetic Braking Efficiency}
Theoretical and simulation studies typically assume alignment between the angular momentum direction of molecular clouds and their magnetic field orientation, implying that a nearly perpendicular misalignment ($\sim$90$^{\circ}$) is almost impossible \citep{Basu1994magnetic, Ouyed1997numerica, Allen2003collapse, Mellon2008magnetic, Machida2011effect, Bate2014collapse, Wurster2018thecollapse}. This convention stems from classical magnetic braking theory. Magnetic braking describes the effect where rapidly rotating gas near the center of a magnetized collapsing core remains magnetically coupled to slower-rotating outer gas. The differential rotation distorts the magnetic field lines, slowing the rotation of the inner gas while transferring angular momentum outward. Classical theory \citep{Mouschovias1980magnetic, Mouschovias1985angualr} predicts that angular momentum parallel to the magnetic field dissipates over a longer timescale than the perpendicular component. Thus, even with an initial misalignment, the perpendicular angular momentum dissipates rapidly, leading to alignment of the rotational axis with the magnetic field. 

However, this conclusion relies on assumptions applicable only to non-collapsing cores where gravity is not dominant. During gravitational collapse, infalling gas drags the field lines into a fan-shaped configuration, distinct from the uniform field assumed classically. Recent analyses and simulations \citep{Joos2012protostellar, Tsukamoto2018does} indicate that magnetic braking efficiency is significantly enhanced in the alignment case with this fan-shaped configuration, exceeding that in the perpendicular case. In other words, within collapsing cores, the perpendicular component of angular momentum is dissipated more slowly than the parallel component of angular momentum. The key reason for this apparent contradiction with classical theory is that the fan-shaped magnetic fields possess longer moment arms, sweeping more material and thus transferring angular momentum more efficiently. Based on this theory, the relationship between the initial magnetic field direction and magnetic braking efficiency has been investigated. Both ideal and non-ideal MHD simulations show that initial misalignment significantly reduces braking efficiency and subsequently influences the magnetic field configuration \citep{Hennebelle2009disk, Joos2012protostellar, Li2013does, Zhao2020formation}. Specifically, simulations indicate that as the initial misalignment angle increases, cores tend to form larger disks, weaker outflows, and exhibit lower magnetic braking efficiency. While some studies report contradictory findings \citep{Matsumoto2004directions, Tsukamoto2018does}, this discrepancy may arise because those simulations focused solely on the early accretion phase, prior to the field being significantly dragged into a fan-shaped configuration \citep{Hirano2020theeffect, Tsukamoto2023therole}.

\begin{figure}[htbp]
	\centering  
    \includegraphics[width=0.8\linewidth]{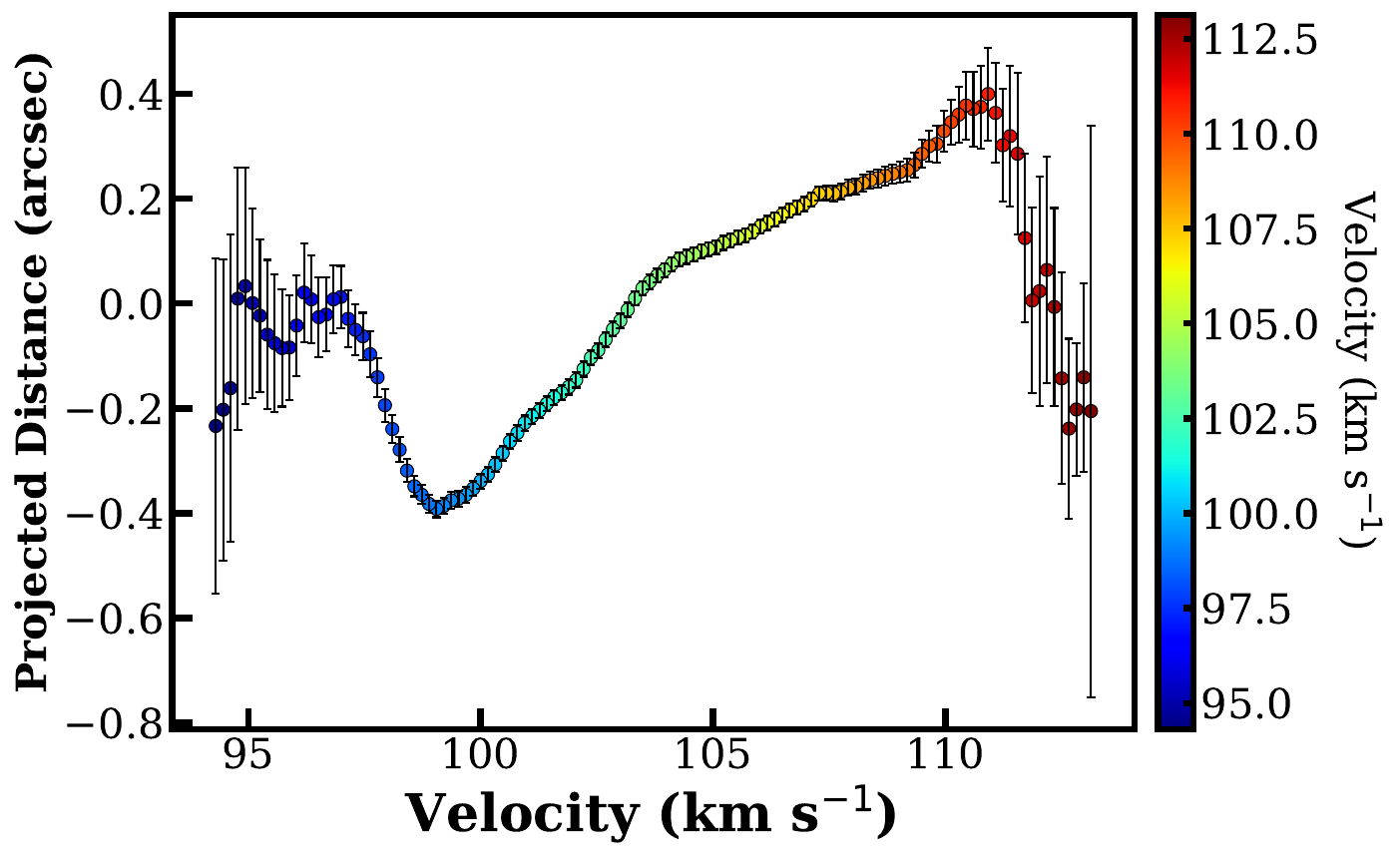}
    \vspace{0.5cm}

    \begin{minipage}[b]{0.48\linewidth}
        \centering
        \includegraphics[width=0.9\linewidth]{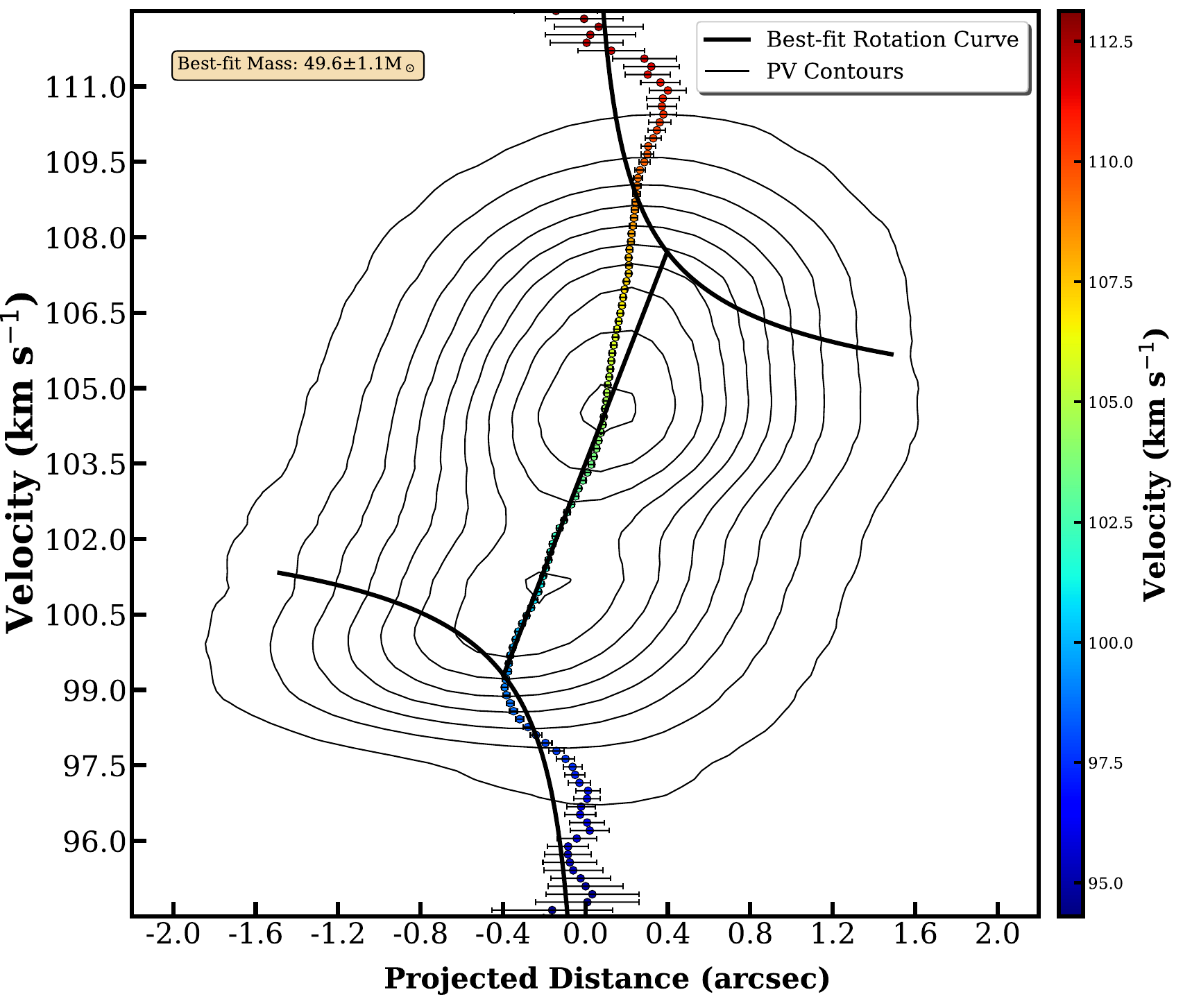}
    \end{minipage}
    \hfill
    \begin{minipage}[b]{0.48\linewidth}
        \centering
        \includegraphics[width=0.9\linewidth]{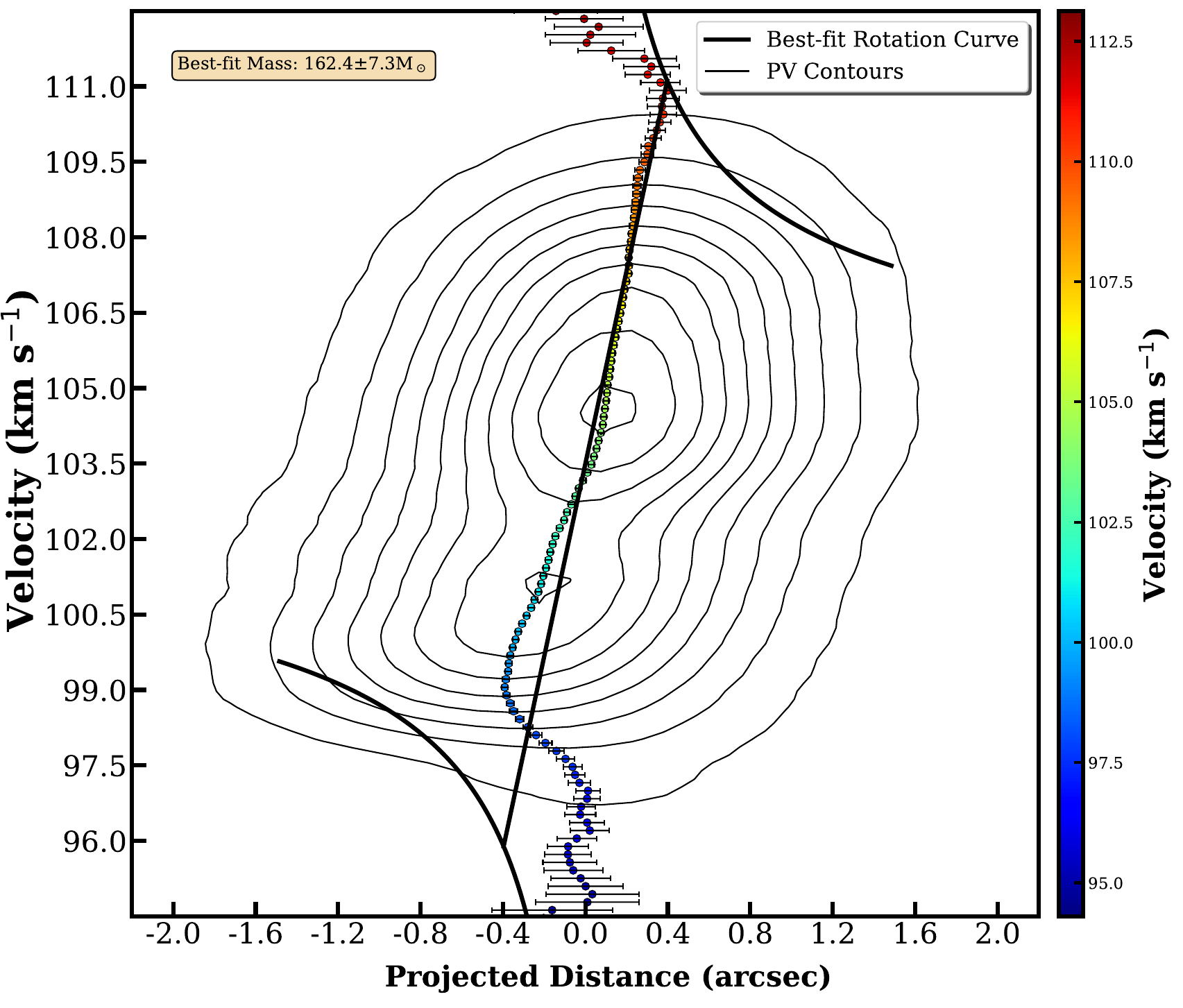}
    \end{minipage}

	\caption{Top panel: Rotation curves for CH$_3$OH($10_{2,9}-9_{3,9}$) in IRAS 18360. The projected distance is calculated as the distance from the velocity centroid point to the best-fitted line passing through the maximum flux position of this transition. Bottom panel: The left and right plots show the best fits to the part of rotation curves for the blue-shifted lobe (below 103 km s$^{-1}$) and the red-shifted lobe (above 104 km s$^{-1}$), respectively. The best-fit enclosed masses are 49.6$\pm$1.1$M_\odot$ for the blue lobe and 162.4$\pm$7.3$M_\odot$ for the red lobe. The black lines represent the resulting fits for each lobe. The contours show the position-velocity diagram along the position angle of 127$^\circ$. The contour levels begin at 3$\sigma$ (60 mJy beam$^{-1}$) and increase to a maximum value of 2.33 Jy beam$^{-1}$ in 10 linearly spaced levels.}
	\label{fig: centriod_fitting}
\end{figure}

Regarding the source IRAS18360, a high-density core exhibiting convincing evidence of collapse \citep{Qiu2012forming}, the uniform density conditions assumed by the classical theory may not hold. Consequently, it is reasonable to explain the nearly 90-degree misalignment observed in our ALMA data with more realistic theories. Furthermore, comparison with the JCMT BISTRO observations of the source reveals a similar magnetic field orientation on larger scales ($\sim$0.5 pc, private communication with the BISTRO consortium). This suggests that the initial magnetic field in this region was also perpendicular to the angular momentum direction. Additionally, several properties of this source match simulation results in which the initial magnetic field is perpendicular to the angular momentum. These simulations predict that due to reduced magnetic braking efficiency, more angular momentum can be retained at the central region, enabling the formation of a larger rotation-supported disk. This is consistent with the large rotational energy and radius of the rotating structures (Figure \ref{fig: centriod_fitting}). Furthermore, the simulations predict increased cloud fragmentation, consistent with the interpretation of condensation MM2 as an envelope fragment. Finally, in this context, the strength of the magnetic field is not expected to be very high, since the efficiency of magnetic braking is positively correlated to the field strength, and thus again the DCF estimate of the field strength should be taken as an upper limit.

\section{Summary and conclusions}
We investigate the magnetic field properties in the high-mass star-forming region IRAS 18360-0537. Our analysis is based on the high-resolution ALMA Band 6 polarization observations, combined with archival VLA observations of NH$_3$(1,1) and NH$_3$(2,2) lines \citep{Lu2014very}, and SMA observations of 1.3 mm dust continuum \citep{Qiu2012forming} and polarization emission \citep{Zhang2014magnetic}. The main findings are as follows:

1. The linearly polarized 1.3 mm continuum emission reveals a highly ordered, hourglass-shaped magnetic field projected on the plane of the sky. The field is predominantly oriented northwest-southeast, with an average position angle of $\bar{\phi}_{PA}=127^\circ$. Intriguingly, the mean field orientation is nearly perpendicular to the outflow/rotation axis, which is challenging to interpret based on magnetic-dominated collapse models for a $\sim$10,000 AU core.

2. We investigate possible scenarios underlying this specific magnetic field morphology and the $\sim$90$^{\circ}$ misalignment between the mean field orientation and the outflow/rotation axis. The correlation between the hourglass magnetic field edge and the outflow cavity walls indicates that the magnetic field is likely shaped by the outflow. Furthermore, we speculate that the flattened field lines are influenced by rotation and gas accretion. Quantitatively, we find that the importance of rotation is comparable to that of the magnetic field, even when adopting a possibly overestimated field strength. We further estimate the rotational-to-gravitational energy ratio comparison and carry out the centroid velocity analysis, both of which reveal a strong rotational structure. Moreover, we develop a simple model showing that gas accretion onto the central high-mass protostar could also play an important role in dragging the field lines into the observed orientation.

Overall, our observations of a high-mass star-forming region showcase a well-shaped, but misaligned (with respect to the rotation axis) hourglass magnetic field that is not a consequence of magnetically-dominated collapse, but instead heavily influenced by bulk gas motions, including outflow, rotation, and accretion. The misalignment could lead to reduced magnetic braking and the formation of large disks. 

\begin{acknowledgments}
Acknowledgments: This work is supported by National Natural Science Foundation of China (NSFC) grant No. 12425304 and by the National Key R\&D Program of China with Nos. 2023YFA1608204 and 2022YFA1603103, and the grant from the China Manned Space Project.
\end{acknowledgments}

\appendix
\section{The estimation of magnetic field strength using DCF method}\label{app:dcf}
Davis–Chandrasekhar–Fermi analysis (DCF; \citealp{Davis1951polarization, Chandrasekhar1953magnetic}) is commonly used to estimate the magnetic field strength. The DCF method assumes that turbulent kinetic energy equals turbulent magnetic energy, suggesting that distortions of the magnetic field lines are caused by turbulence and such distortions are relatively small (<25$^\circ$, \citealp{Ostriker2001density}). However, the application of the DCF method has certain limitations. Specifically, the assumption of energy equipartition fails when high-density cores are dominated by gravity. Similarly, when the shape of the magnetic field within a region is no longer influenced by motion disturbances other than magnetic fields and turbulence, the assumption of energy equipartition also becomes invalid \citep{Liu2022magnetic}. The perpendicular orientation of the magnetic field relative to the outflow direction indicates that the magnetic field might be influenced by the gas. As a result, the basic assumptions required for the DCF method may not be satisfied. Nevertheless, the DCF method can still be used to provide an estimate of the magnetic field strength, serving as a reference for determining the relative magnitudes of the magnetic field and other physical quantities.

The strength of the ordered component of the magnetic field in the plane of the sky is estimated using the approach proposed by \citet{Crutcher2004scuba}:

\begin{equation}\label{DCFequation}
	B_{pos}=Q\sqrt{4\pi\rho}\dfrac{\delta v_{turb}}{\delta \phi}\approx 10.9 \left[\dfrac{n(H_2)}{10^6 cm^{-3}}\right]^{1/2} \left(\dfrac{\delta v_{turb}}{1 km s^{-1}}\right) \left( \dfrac{\delta \phi}{1^\circ}\right)^{-1} mG
\end{equation}

In this equation, $Q$ is a dimensionless correction parameter. The value of $Q\sim 0.5$ for strong magnetic fields ($\delta \phi \le 25^{\circ}$) is commonly used in previous studies. However, this factor is based on simulations conducted on low-density molecular clouds with scales larger than parsecs \citet{Ostriker2001density}. Here, we adopt a correction factor of $Q \sim 0.28$ derived from a recent simulation for strong field models at clump and core scales \citep{Liu2022magnetic}. The gas volume density is denoted by $\rho=\mu m_{H} n(H_2)$, where $\mu=2.86$ represents the mean molecular weight (\citealp{Kirk2013first}). 

\subsection{The angular dispersion}\label{app:angle_discusssion}
The contribution of the angular dispersion from the large-scale field structure should be removed. Considering the clear hourglass shape of the observed magnetic field, we preferred to fit the large-scale field morphology using analytical hourglass models, as had been done in previous studies (e.g., \citealp{Girart2006magnetic, Rao2009iras, Qiu2014submillimeter, Stephens2013magnetic}). To provide a simple representation of the field geometry, we adopted a parabolic function model as proposed by \citet{Girart2006magnetic}, with the equation of $y=g(1+Cx^2)$. In this equation, the $x$-axis represents the mean magnetic field direction, and the $y$-axis is the corresponding orthogonal coordinates through the symmetry center. We found the minimized $\chi^2$ solutions that the field symmetry center is located at $(R.A., decl.)_{J2000} = (18^h38^m47^{s}.4,-5^{\circ}35^{\prime}4^{\prime\prime}.0)$, the symmetry axis angle is $\bar{\theta}_{B}=127^{\circ}$ and the shape controlling parameter is $C=0.009$ in the unit of pixel size (0.15{$^{\prime\prime}$). The position angle residuals between the observed data and models are defined as $\phi = \theta_B - \theta_{parab}$. Figure \ref{fig: Model_fitting}(a) shows the corresponding histograms of residual angles of the fitted field orientations. The standard deviation of this residuals is $\delta \phi_{parab}=5.8^{\circ}$. Therefore, the corrected dispersion comes to $\delta \phi_{cor,parab}=5.6^{\circ}$, considering the angle measurement uncertainty of $\bar{\sigma}_{\phi}=1.4^{\circ}$  based on S/N of the polarized emission.\par

Moreover, we further adopted a meticulous fitting based on a more realistic model from the theoretical solutions of Green’s function \citep{Ewertowski2013mathenatical}. The expressions of the symmetric magnetic field $B(x,y)$ satisfy the following equations:

\begin{equation}
    B_x=\sum\limits_{m=1}^{\infty} k_m \sqrt{\lambda_m} J_1(\sqrt{\lambda_m}r)\left[ erfc\left( \dfrac{\sqrt{\lambda_m} h}{2}-\dfrac{y}{h} \right) e^{-\sqrt{\lambda_m}y} - erfc\left( \dfrac{\sqrt{\lambda_m} h}{2}+\dfrac{y}{h} \right) e^{\sqrt{\lambda_m}y} \right]
\end{equation}	
\begin{equation}	
    B_y=\sum\limits_{m=1}^{\infty} k_m \sqrt{\lambda_m} J_0(\sqrt{\lambda_m}x)\left[ erfc\left( \dfrac{\sqrt{\lambda_m} h}{2}+\dfrac{x}{h} \right) e^{\sqrt{\lambda_m}y} + erfc\left( \dfrac{\sqrt{\lambda_m} h}{2}-\dfrac{z}{h} \right) e^{-\sqrt{\lambda_m}y} \right]+B_0 
\end{equation}
	
where the $y$-axis represents the mean magnetic field direction and the $x$-axis is the corresponding orthogonal coordinates through the symmetry center. $J_0$ and $J_1$ are Bessel functions of the first kind of order 0 and 1, respectively. $\lambda_m=({a_{m,1}}/{R})^2$ where $a_{m,1}$ is the $m$ root of $J_1(x)$. $R$ and $h$ represent the radius and thickness of the disk used in this model. $k_m$ are coefficients that are treated as unknown fitting parameters. We set $B_0=1$, representing the unit strength of the background magnetic field. We minimize the $\chi^2$ of the position angle residual obtained from $\phi=\theta_B-\theta_{Emodel}=\theta_B-arctan\left({B_z}/{B_r} \right) - 90^\circ$. To reduce the free parameters, we fixed the symmetry center and axis orientation to be the same as in the parabolic model. Only the first three terms in the series need to be considered \citep{Ewertowski2013mathenatical}, resulting in flexible parameters to $R, h, k_1,k_2,k_3$. The least $\chi^2$ fitting parameters are $R=22.5$, $h=7.20$, $k_1=48,k_2=77,k_3=314$, also in the unit of pixel size. Figure \ref{fig: Model_fitting}(b) shows the histogram of residual angles derived by this model, yielding a standard deviation of $\delta \phi_{Emodel}=4.9^{\circ}$ and the corrected dispersion of $\delta \phi_{cor, Emodel}=4.7^{\circ}$. The residual angles of the second model revealed a better Gaussian-like profile and a narrower distribution, which indicates a more realistic fit to the large-scale field lines. Therefore, we adopted the angular dispersion $\delta \phi_{cor}=4.7^{\circ}$ as the final result for the model fitting method.\par

The structure function method is another possible approach to separate the turbulent component from the large-scale ordered field, initially proposed by \citet{FalcetaGonalves2008studies} and \citet{Hildebrand2009dispersion}. The advantage of this method is that it allows us to obtain the magnetic field angle dispersion without relying on any specific model. The structure function is defined as the average of the difference between polarization angles for all pairs of points with vector separation:
\begin{equation}
    \langle\delta \Phi(l)^2\rangle^{1/2}=\{\dfrac{1}{N(l)}\sum_{i=1}^{N(l)}[\Phi(\textbf{x})-\Phi(\textbf{x+l})]^2\}^{1/2}
\end{equation}

where $\Phi(\textbf{x})$ and $\Phi(\textbf{x+l})$ are the polarization angles at locations of $\textbf{x}$ and $\textbf{x+l}$ respectively, and $N(l)$ is the number of pair of points having same separation $l=|\textbf{l}|$. Assuming that the magnetic field consists of a large-scale ordered field $B_0$ and a turbulent component $\delta B$ with the correlation lengths $d$ and $\delta$. When $\delta<l\ll d$, the higher-order terms of the Taylor expansion can be canceled out, and the square of the structure function is approximately expanded to $\langle\delta \Phi(l)^2\rangle=b^2+m^2 l^2+\sigma^2_M(l)$, as a quadratic function of separation. The parameter $b^2$ represents the turbulent contribution, $m^2l^2$ represents the large-scale structured field contribution, and $\sigma^2_M(l)$ represents the contribution of measurement uncertainties. The estimated angular dispersion and the ratio of the turbulent to large-scale magnetic field strength are expressed as
\begin{equation}
    \delta \phi \approx \dfrac{\langle \delta B^2\rangle^{1/2}}{B_0}=\dfrac{b}{\sqrt{2-b^2}}
\end{equation}
	
Figure \ref{fig: SFmap} shows the measured angle difference as a function of the separation $l$. At scales of 2.2$^{\prime\prime}$-3.6$^{\prime\prime}$, the measured structure function has an opposite trend with distance, possibly because of incomplete sampling caused by a few polarization angle pairs. The fit to the structure function was carried out for $1.1^{\prime\prime} < l< 2.2^{\prime\prime}$ to avoid the influence below the synthesized beam. We find the turbulent contribution $b \approx 13.0^{\circ} \pm2.5^{\circ}$ ($0.22 \pm 0.04 rad$), equivalent to angle dispersion $\delta \phi= 9.2^{\circ} \pm 1.8^{\circ}$ and the turbulent to large-scale magnetic field strength ratio $ \langle \delta B^2\rangle^{1/2} /B_0=0.16\pm 0.03$. This value is approximately twice that obtained from the model fitting method, but it remains significantly lower than the standard deviation of the field position angle $\sigma_{\theta_B}=18^{\circ}$. We have also applied the angle dispersion function approach (ADF, \citealp{Houde2016dispersion}), while the turbulent-to-total field ratio diverges with the scale. This might be due to the significant change in magnetic field orientations at smaller scales.

It is important to stress that the angular dispersion obtained through those conventional methods is uncertain because the observed hourglass-shaped magnetic field does not entirely match the expectations of strong magnetic field theories. From a purely statistical perspective, determining which method is more reliable depends on whether the contribution of the ordered field to angular dispersion can be accurately separated without overfitting. However, it is not possible to definitively claim that one method is more reliable than the other. Therefore, we have presented all the results above.

\subsection{Temperature and volume density}\label{app:temp_discussion}
The mass density is calculated based on the gas mass of 80$M_{\odot}$ from SMA observations \citep{Qiu2012forming}. We further estimate the gas mass from the dust continuum emissions. With the assumption of optical thinness and gas-to-dust mass ratio of $\Lambda=100$, the total gas can be obtained by the following equation: 
\begin{equation}
    M=\dfrac{\Lambda F_{\nu} D^2}{\kappa_{\nu}B_{\nu}(T)}
\end{equation}
where $\Lambda$ is the gas-to-dust mass ratio, $F_\nu$ is the source flux density, $D$ is the distance to the source, $\kappa_{\nu}=(\nu/1 THz)^\beta$ is the dust opacity, and $B_{\nu}$ is the Planck function at the dust temperature $T$. We used the dust opacity of $\kappa_{\nu}=2.9cm^2 g^{-1}$ with the opacity index of $\beta=0.82$ (\citealp{Qiu2012forming}). 

Although our ALMA observations did not give precise dust temperature measurements, we can estimate the upper and lower limits of dust temperature using data from two typical molecules. Methanol is one of the hot gas tracers. The rotation temperature map (see Figure \ref{fig: ch3oh_temp_map}) is generated using the rotational diagram method \citep{Goldsmith1999population} based on six not blended CH$_3$OH lines (Table \ref{table: ch3oh_used_temp}). Under the assumption that the gas is in LTE and the lines are optically thin, the integrated line intensity $\int T_{b} dV$, rotational temperature $T_R$, and the molecular gas column density are related by\citep{Liu2020chemistry}:
\begin{equation}
    ln\left( \dfrac{3k\int T_{b}dV}{8\pi^3\mu^2\nu S}\right) = ln\left( \dfrac{N_u}{g_u}\right)=ln{N}-ln{Z} -\dfrac{E_u}{kT_{R}}
\end{equation}
Where $N_u$ is the upper state column density and $g_u$ is the corresponding state degeneracy. $Z$ is the rotational partition function. $S$ is the line strength of the transition. The average temperature of the center region is $T_{R, CH_3OH}=160 \pm 50K$ (see Figure \ref{fig: ch3oh_temp_map}). The minimum gas mass limit derived from this rotational temperature map is $M_{min}=39M_{\odot}$. The average volume density of $n(H_2)_{min}=4.9\times 10^{5} cm^{-3}$. Ammonia is a suitable cold-dense gas tracer due to its presence in the gas phase within highly dense environments \citep{Beuther2002massive} and cold molecular cores \citep{Tafalla2002systematic}, where other molecules like CO are partially frozen onto dust grains. The NH$_3$(1,1) transition consists of a main component ($\Delta$F = 0) and four satellite components ($\Delta$F = $\pm$1), evenly located on both sides of the main component\citep{Ho1983ammonia}. The VLA data covers the main component and two satellite components (\citealp{Lu2014very}). We fitted this triple peak structure of the NH$_3$(1,1) lines using the \textit{CLASS}, from the software package GILDAS\footnote{Available at \url{https://www.iram.fr/IRAMFR/GILDAS}}. The fitted velocity dispersion and the optical depth of the primary peak are $\sigma_{v11} = 2.0 \pm 0.1$ km s$^{-1}$ and $\tau_{m}(1,1) = 1.2 \pm 0.3$, respectively. The velocity dispersion and the optical depth of NH$_3$(2,2) transition main component are $\sigma_{v22} = 2.3 \pm 0.2$ km s$^{-1}$ and $\tau_{m}(2,2) = 1.4 \pm 0.4$, respectively. The rotational temperature of NH$_3$ can be expressed by \citep{Ho1983ammonia}:

\begin{equation}
    T_R (2,2;1,1) = -41.5\div ln\left[\dfrac{-0.282}{\tau_m (1,1)} ln \left\{1 - \dfrac{T_a(2,2,m)}{T_a(1,1,m)} \times (1 -e^{-\tau_m (1,1))} \right\} \right]
\end{equation}

Where ${T_a(1,1,m)}$ and ${T_a(2,2,m)}$ represent the antenna temperatures for the primary peaks of NH$_3$(1,1) and NH$_3$(2,2) lines, respectively. $\tau_{m}(1,1)$ is the primary peak optical depth of NH$_3$(1,1). The rotational temperature of ammonia molecules is estimated to be $T_{R, NH_3} = 34.5K$. The relationship between rotational temperature and kinetic temperature can be described by the following equation \citep{Walmsley1983ammonia}:
\begin{equation}
    T_R = \dfrac{T_K}{1 + \dfrac{T_K}{41.5} ln\left[1 + 0.6 exp\left(-\dfrac{15.7}{T_K}\right)\right] }
\end{equation}

We estimated the kinetic temperature of $T_{K, NH_3}=55K$, and considered it as the lower limit of the average dust temperature. The maximum values for mass and volume density as $M_{max}=105M_{\odot}$ and $n(H2)_{max}=1.3\times 10^{6} cm^{-3}$, respectively.

\begin{figure}[htbp]
	\centering
	\includegraphics[width=0.8\linewidth]{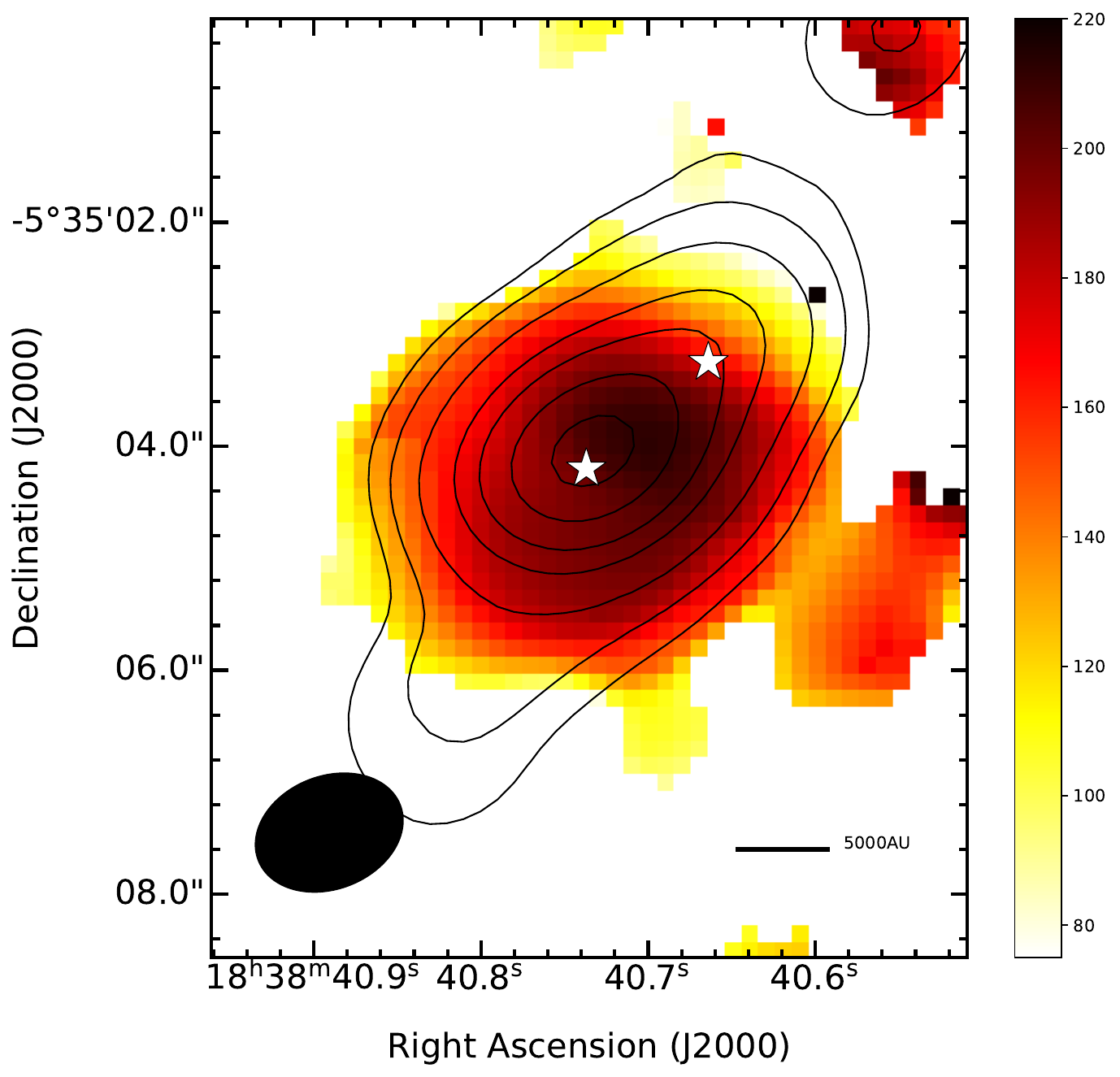}
	\caption{The rotation temperature map for CH$_3$OH. Only the locations with significant measurements in integrated flux (above $3\sigma$) in all of the involved transitions are included in the map. The continuum contours adopt the same levels as Figure \ref{fig:pibmap}(b) but begin at the 12$\sigma$.}
	\label{fig: ch3oh_temp_map}
\end{figure}

\begin{table}[htbp] 
	\caption{Summary of the used CH$_3$OH transitions} 
	\label{table: ch3oh_used_temp}
	\centering
	\begin{tabular}{cccl}
		\hline 
		Molecule & Transition & Frequency[GHz] & $E_{up}/k$ [K] \\ 
		\hline 		
		CH$_3$OH  &  $5_1 - 4_2$              &  216.9455 & 55.87  \\ 
                  &  $6_{1,5} - 7_{2,6}$      &  217.2992 & 373.93 \\
                  &  $20_1 - 20_0$            &  217.8865 & 508.38 \\
		         &  $4_2 - 3_1$              &  218.4401 & 45.46  \\ 		
		         &  $10_{-3,8} - 11_{-2,10}$ &  232.9458 & 190.37 \\
		         &  $18_{3,15} - 17_{4,14} $ &  233.7958 & 446.58 \\
		\hline     	            
	\end{tabular} 
\end{table}

\subsection{The turbulence velocity}\label{app:turb_discussion}
The turbulent velocity varies with the observation scale (\citealp{Larson1981interstellar, FalcetaGoncalves2014turbulence}). In most studies estimating field strength via the DCF method, the gas turbulent velocity is directly derived from spectral linewidth (\citealp{Lai2001interferometric, Heiles2005magnetic, Rao2009iras, PlanckCollaboration2016planckintermediate, Pattle2023magnetic}). Correspondingly, the linewidth of the high excited state CH$_3$OH ($10_{2,9}-9_{3,9}$) lines reaches 2.7 km s$^{-1}$. However, the observed velocity dispersion is often influenced by bulk motions. Several works qualitatively discussed such turbulence overestimation affected by rotation, accretion, infall motions, and outflows (\citealp{Santos2016magnetically, Aso2021multi}), but these discussions did not provide quantitative values. Because the thermal velocity is negligible, we decompose the dispersion into components of turbulence and bulk motions, that is $\sigma_{v}^2 = \sigma_{turb}^2 +\sigma_{bm}^2$. According to \citet{Yue2021resolution}, the bulk motion component can be partially estimated from the centroid velocities of the magnetic field detected region via $\sigma_{bm}^2=\sum_{i=1}^{N}(v_i-\bar{v})^2\ J_i/J$. Where $v_i$ is the centroid velocities at pixel $i$ and $\bar{v}$ is the average centroid velocities of the statistic region; $J_i$ is the spectral line flux density at pixel $i$ and $J$ is the total flux density. After deducting the bulk motions, the velocity dispersion is reduced to 2.0 km s$^{-1}$. This value is close to the velocity dispersion obtained from the $NH_3$ lines. However, we would like to emphasize that the turbulence velocity dispersion obtained through this method may still be prone to overestimation. For example, rotational motions significantly influence the velocity map, whereas infall and outflow motions have a comparatively lesser impact. Moreover, the bulk motions with a physical scale smaller than the resolution are unable to be counted. To investigate the impact of bulk motions on turbulence measurement more completely, we performed a radiative transfer simulation with an envelope-disk model using the RADMC-3D package\footnote{Available at \url{https://www.ita.uni-heidelberg.de/~dullemond/software/radmc-3d/}} \citep{Dullemond2012RADMC-3D}. The best-fitted parameters for this model are a protostar mass of 60$M_{\odot}$ and a turbulence velocity of 1.4$km/s$ (see details, \citealp{Mo2023exploring}). We note that a single protostar with a mass of 60$M_{\odot}$ would result in ionized hydrogen radiation regions. Regardless, we mainly considered the protostar mass in terms of dynamics in the model. Therefore, the protostar mass in the model may be contributed by several smaller protostars. Nevertheless, we have effectively eliminated the influence of bulk motions, particularly infall motions, which cannot be deduced from velocity maps. Therefore, we ultimately derived a turbulence velocity of 1.4$km/s$.\par

\subsection{Comparison of the magnetic field, gravity, and turbulence}\label{app:compBGT}
The values of several variables that significantly influence the estimation of magnetic field strength are discussed in the preceding section. The total gas mass is 80$M_{\odot}$ from the SMA observations \citep{Qiu2012forming} for the detected magnetic field region of $2.5^{\prime\prime} \times 1.95^{\prime\prime}$. We assume that the line of sight axis aligns with the minor axis of the elliptical region. The volume density is computed to be $1.0\times 10^{6} cm^{-3}$. The upper and lower limits of the gas mass result in a volume density error bar ranging from 0.7 to 1.15 times. The turbulence velocity is 1.4$km\ s^{-1}$ and the angular dispersion of the field lines is calculated to be $\delta \phi=4.6^\circ$ and $\delta \phi=9.2^\circ$ using the fitting hourglass model and structure function method, respectively. In summary, the magnetic field strength is estimated to be $B_{hmodel}=3.2mG$ and $B_{sf}=1.7mG$ by equation \ref{DCFequation}.

We compared the magnetic field with gravity through the mass-to-flux ratio. The observed mass-to-flux ratio, expressed in units of the critical value $(M/\Phi)_{cirt}=1/2\pi\sqrt{G}$ \citep{Nakano1978gravitational}, is defined as: $\lambda_{obs}= f{(M/\Phi)_{obs}}/{(M/\Phi)_{crit}}= 7.6\times 10^{-21} f{N(H_2)}/{B}$ \citep{Crutcher2004scuba}. $N(H_2)$ is measured in $cm^{-2}$ and $B$ is the field strengths in $\mu$G. The correction factor $f$ is determined by the geometry of the magnetic field. In our observations, we have found that the oblate spheroid core is compressed parallel to the magnetic field, with a suggested value of 3/4 for $f$. By contrast, a perpendicular field is associated with an optimal factor of 1/3, whereas a spherical core corresponds to a factor of 1/2 \citep{Crutcher2004scuba}. The mass-to-flux ratio within is computed as 0.88 for the model fitting method and 1.74 for the structure function method. Table \ref{table:compre_G_B_T} shows the comparison of the magnetic field, gravity, and turbulence. The turbulent-to-magnetic energy ratio is estimated to $\beta_{trub}\sim 0.1-0.6$ for these two methods. Both results suggest that the magnetic field dominates over turbulence; however, the difference between them lies in the comparison of the magnetic field to gravity. The former subcritical value indicates that the magnetic field is either slightly dominant or comparable to gravity, while the latter suggests that the magnetic field is insufficient to prevent gravitational collapse. Moreover, a comprehensive statistical analysis of all DCF detections reveals that the contribution from the ordered field structure can overestimate the angular dispersion by an average factor of approximately 2.5 (Liu et al. 2022). This average value is more consistent with the results obtained from the structure function method. Moreover, from an observational perspective, previous studies have already detected an infalling envelope through CN(2-1) inverse P-Cygni profiles \citep{Qiu2012forming}. Additionally, blue asymmetries are detected in the HCO(3-2) and CH$_3$OH lines from our observations, which provide further evidence supporting gas infalling. In this context, a weaker magnetic field strength is more physically realistic. Nonetheless, we place greater confidence in the latter result, or alternatively, we believe that both results may be significantly overestimated. 

The complementary method to check our proposal is the polarization–intensity gradient technique \citep{Koch2012magnetic}. In this method, the orientation of the intensity gradient is assumed to represent the direction of material motions resulting from pressure, magnetic field, and gravity. It is significant to note that this assumption does not rely on a strong magnetic field environment. The ratio between magnetic field force $F_B$ and gravitational force $F_G$ at each detected location can be expressed as $\Sigma_B = {F_B}/{F_G}={sin\psi}\ {sin(90^{\circ}-\delta)}$ \citep{Koch2012magnetic}. The parameters $\psi$ and $\delta$ represent the angular differences between the intensity gradient and local gravity orientations, and between the intensity gradient and magnetic field orientations, respectively. We determine the mean ratio of the same region as the DCF method to be $\left< \Sigma_B \right>=0.4$, indicating that the magnetic field does not dominate gravity (shown in Figure \ref{fig: magnetic2gravity ratio}). This result, to some degree, further supports our proposition that magnetic field lines are probably dragged by gravitational forces and the gas bulk motions.

We argue that neither of the aforementioned methods is entirely accurate, as both focus on magnetic field morphological analysis. The magnetic models rely on a three-dimensional hourglass-shaped magnetic field structure, while the structure function method needs a strong magnetic field assumption. However, as discussed above, the underlying assumptions of them are not entirely reliable. We find that the observed hourglass-shaped magnetic field structure may not represent an ordered, strong magnetic field, but rather a relatively weak magnetic field that is dragged by the gas bulk motions. Therefore, the magnetic field angular dispersion on the sky plane will underestimate the distortion of the three-dimensional magnetic field. Unfortunately, our observations are unable to estimate this distortion using other methods.

Therefore, if a reference value is needed, we can only provide an upper limit for the magnetic field strength. First, the magnetic model-fitting method is the least reliable, as it assumes a highly ordered magnetic field perpendicular to the outflows. Second, directly using the angular dispersion of the magnetic field makes it difficult to quantify the contributions of large-scale magnetic field structures and the projection effects, complicating comparisons with other physical quantities. The structure function method offers a more moderate estimate by eliminating the contributions of large-scale structures without relying on models. Furthermore, when combined with the upper limit for gas mass, we provide an upper limit for magnetic field strength of $B_{pos,u}=1.7 mG$.
 
This case also provides an interesting reminder that the highly ordered hourglass field on the plane of the sky does not necessarily represent a three-dimensional ordered magnetic field. Therefore, it is essential to combine the kinematic information to make a more informed judgment.

\begin{figure}[htbp]
	\centering
    \subfigure[Comparison between the observed and modeled magnetic field orientations]{
        \includegraphics[width=0.44\linewidth]{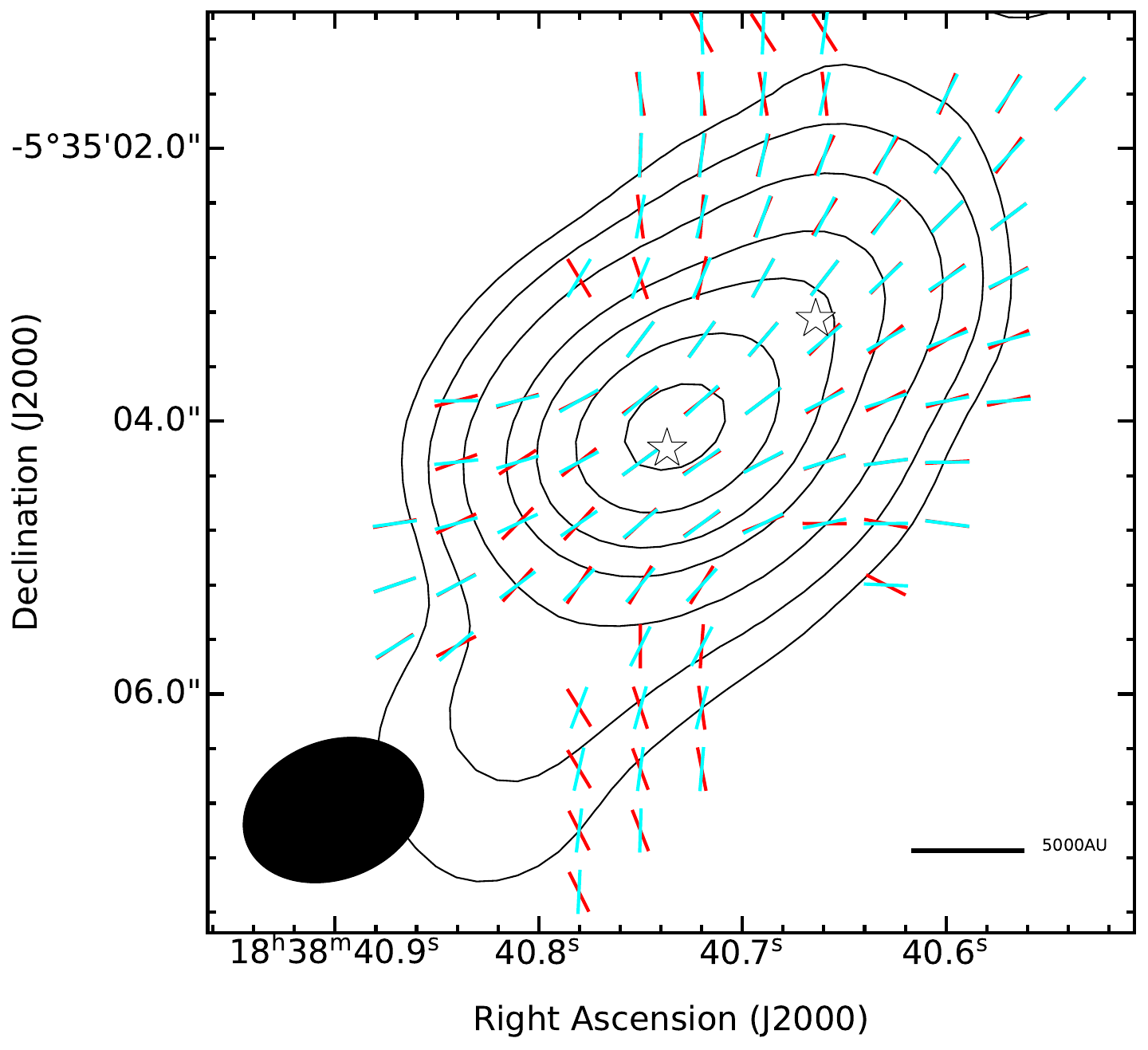}
        \includegraphics[width=0.44\linewidth]{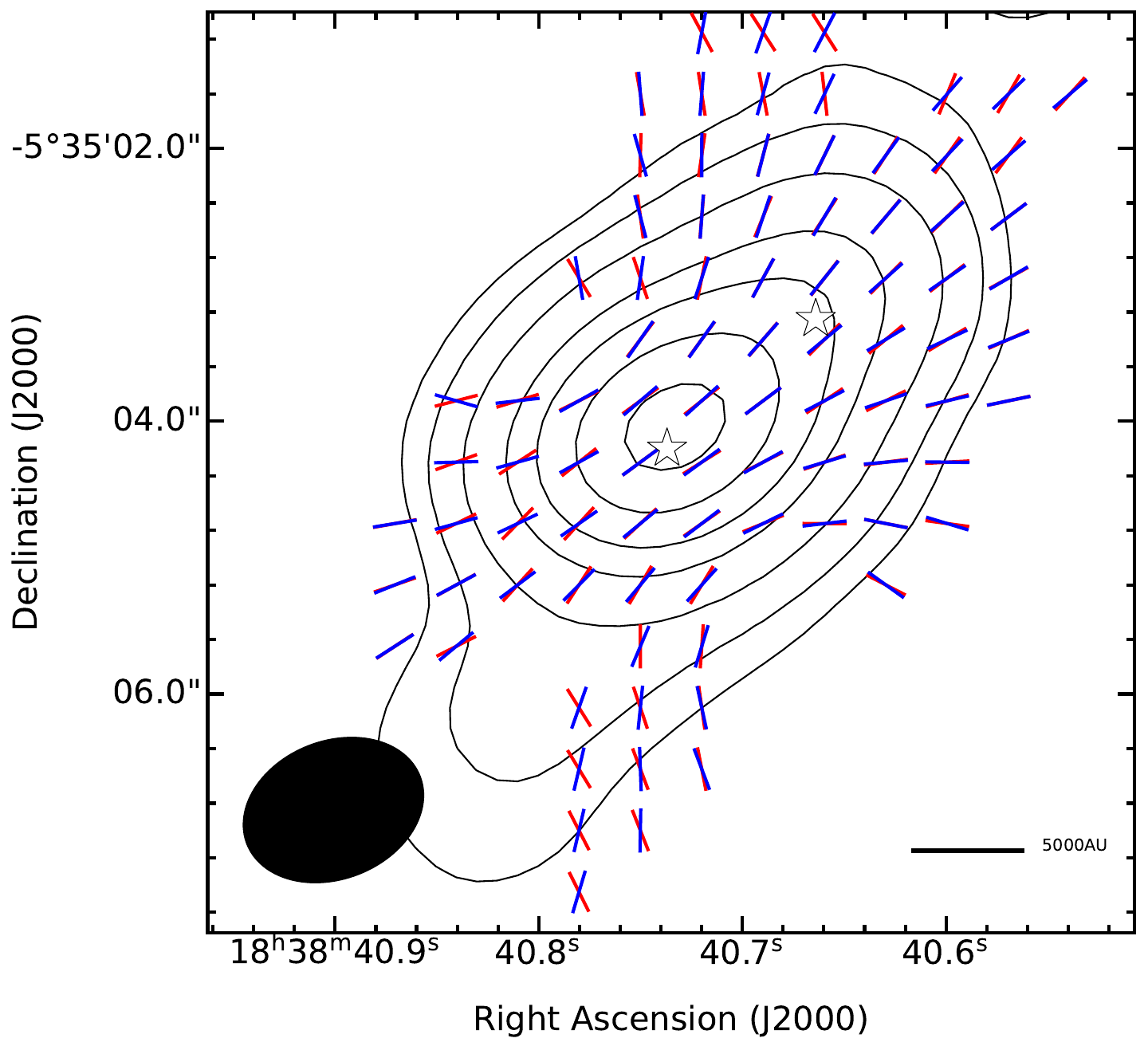}
    }
	\subfigure[Histogram of the polarization angle residuals]{
         \includegraphics[width=0.44\linewidth]{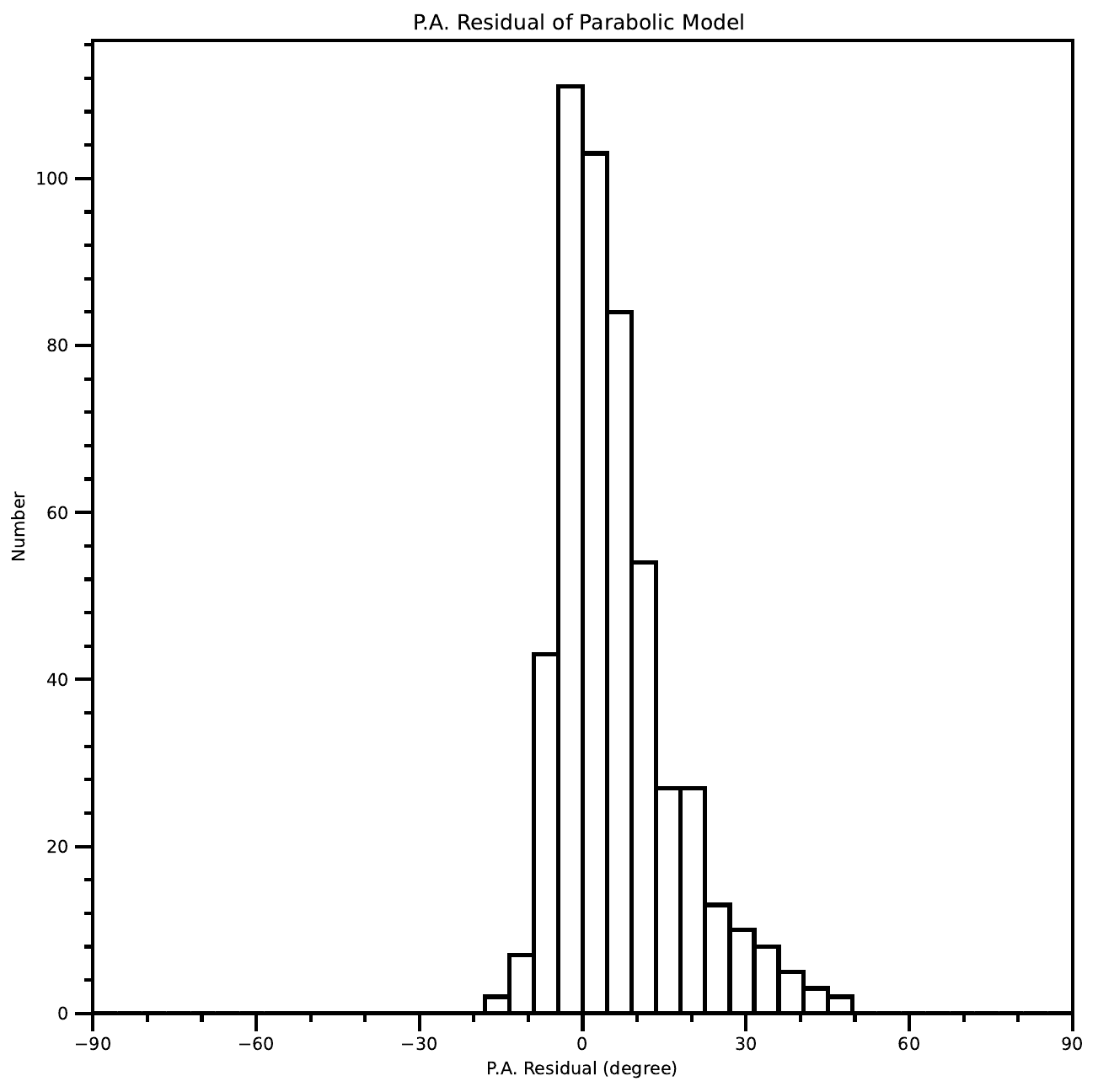}
         \includegraphics[width=0.44\linewidth]{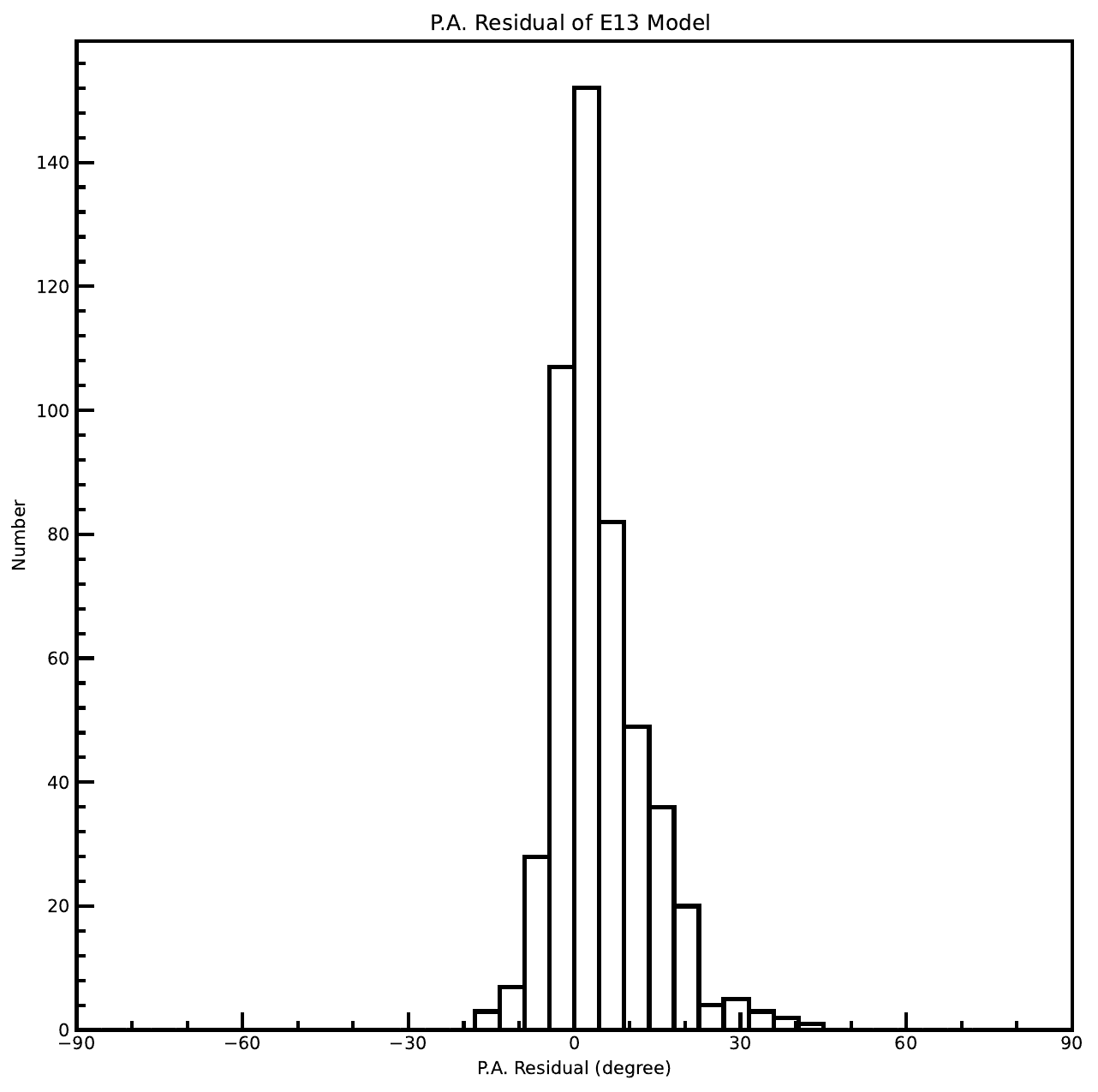}
    }
	\caption{Top panel: Contour map of the total (Stokes I) dust emission and the magnetic field vectors (The black contours are same as Figure \ref{fig: ch3oh_temp_map}) with the best family of functions are shown as B-vectors (Cyan bars: parabolic model; Blue bars: Ewertowski model) at the same position as the measured values. Bottom panel: Position angle (P.A.) distribution of the polarization vectors for the best fit parabolic model and Ewertowski model.}
	\label{fig: Model_fitting}
\end{figure}
\begin{figure}[htbp]
	\centering
	\includegraphics[width=0.8\linewidth]{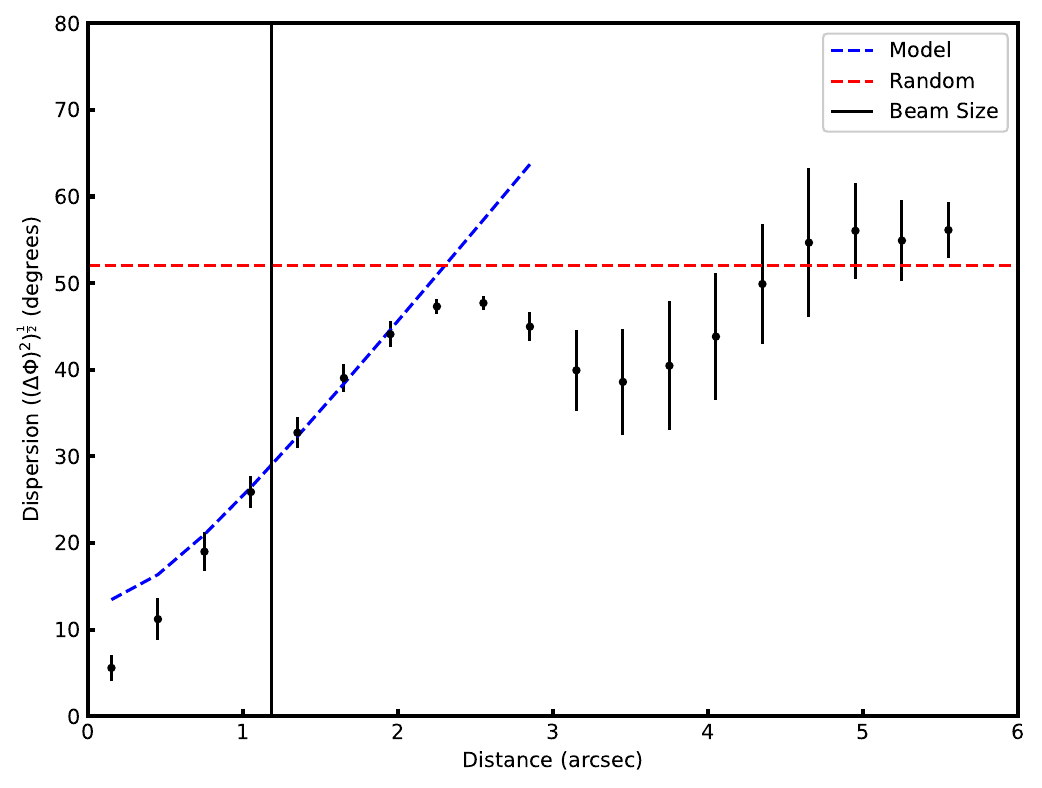}
	\caption{Structure function for a center region of IRAS 18360. The measured angle dispersion is shown in black solid circles with the error bars. The best is shown by the blue dashed line. The vertical dashed line represents the beam size of the observations $\sim 1.1^{\prime\prime}$ and the horizontal red dashed line is the predicted angle dispersion for a random field $\sim 52^{\circ}$.}
	\label{fig: SFmap}
\end{figure}

\begin{figure}[htbp]
	\centering
	\includegraphics[width=0.8\linewidth]{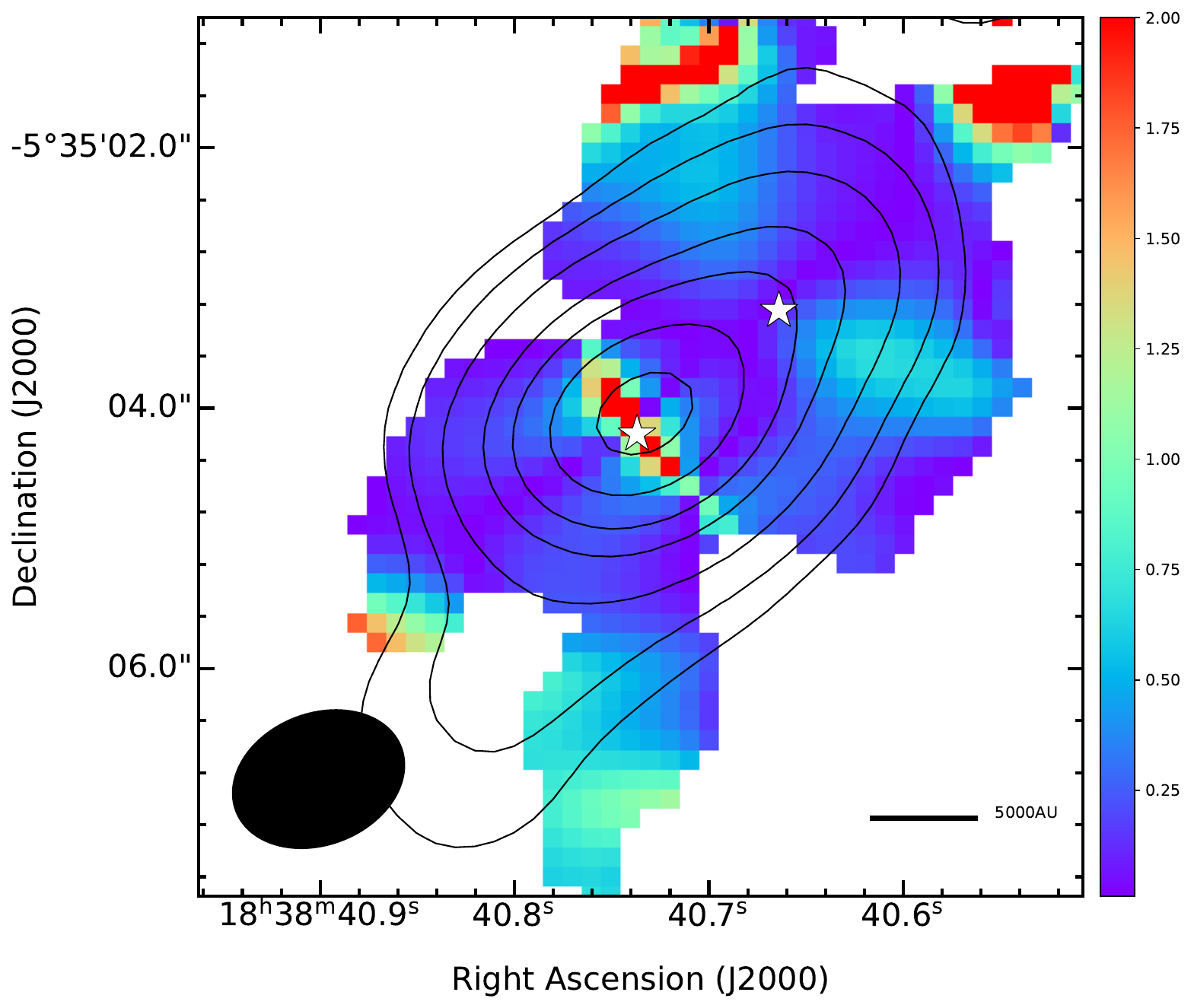}
	\caption{Spatial distribution of the B-field-to-gravity force ratio $\Sigma_{B}$ toward IRAS18360. The contour levels are the same as in Figure \ref{fig: ch3oh_temp_map}. $\Sigma_{B} < 1$ throughout most of the region, except at the condensation MM1 and the edge of the area where the magnetic field is detected.}
	\label{fig: magnetic2gravity ratio}
\end{figure}

\begin{table}[htbp]
    \caption{Comparison of the magnetic field, gravity, and turbulence} 
    \label{table:compre_G_B_T}
    \centering
    \begin{threeparttable}     
	\begin{tabular}{cccll}
		\hline 
		Method & $\lambda$\tnote{1} & $\beta$\tnote{2} \\ 
		\hline 		
		Fit Large scale Field Method & 0.88 &  0.1 \\ 
	    \hline 		
		Structure Function Method & 1.74 &  0.6 \\
		
		\hline     	            
	\end{tabular} 

        \begin{tablenotes}
        \footnotesize
        \item[1] The mass to flux ratio.
        \item[2] The turbulence to magnetic energy ratio.
        \end{tablenotes}
    
    \end{threeparttable} 
\end{table}
\newpage
\bibliography{i18360mag}

\end{document}